\documentclass[twocolumn]{aastex61}

\usepackage{amsmath}

\received{January 23, 2018}
\revised{February 28, 2018}
\accepted{March 8, 2018}
\submitjournal{ApJ}

\shorttitle{RR Lyrae near-IR light curves}
\shortauthors{Hajdu et al.}

\begin{document}

\title{A data-driven study of RR~Lyrae near-IR light curves: principal component analysis, robust fits, and metallicity estimates}

\correspondingauthor{Gergely Hajdu}
\email{ghajdu@astro.puc.cl}

\author[0000-0003-0594-9138]{Gergely Hajdu}
\affiliation{Instituto de Astrof\'isica, Pontificia Universidad Cat\'olica de Chile,
Av. Vicu\~na Mackenna 4860,
782-0436 Macul, Santiago, Chile}
\affiliation{Astronomisches Rechen-Institut, Zentrum f\"ur Astronomie der Universit\"at Heidelberg,
M\"onchhofstr. 12-14,
D-69120 Heidelberg, Germany}
\affiliation{Instituto Milenio de Astrof\'isica,
Santiago, Chile}

\author[0000-0001-7696-8331]{Istv\'an D\'ek\'any}
\affiliation{Astronomisches Rechen-Institut, Zentrum f\"ur Astronomie der Universit\"at Heidelberg,
M\"onchhofstr. 12-14,
D-69120 Heidelberg, Germany}

\author[0000-0003-1404-9270]{M\'arcio Catelan}
\altaffiliation{On sabbatical leave at the Astronomisches Rechen-Institut,
Zentrum f\"ur Astronomie der Universit\"at Heidelberg, Heidelberg, Germany}
\affiliation{Instituto de Astrof\'isica, Pontificia Universidad Cat\'olica de Chile,
Av. Vicu\~na Mackenna 4860,
782-0436 Macul, Santiago, Chile}
\affiliation{Instituto Milenio de Astrof\'isica,
Santiago, Chile}

\author[0000-0002-1891-3794]{Eva K. Grebel}
\affiliation{Astronomisches Rechen-Institut, Zentrum f\"ur Astronomie der Universit\"at Heidelberg,
M\"onchhofstr. 12-14,
D-69120 Heidelberg, Germany}

\author[0000-0001-7901-7689]{Johanna Jurcsik}
\affiliation{Konkoly Observatory, Research Centre for Astronomy and Earth Sciences, Hungarian Academy of Sciences,
PO Box 67, 1525 Budapest, Hungary}


\begin{abstract}

RR~Lyrae variables are widely used tracers of Galactic halo structure
and kinematics, but they can also serve to constrain the distribution of the
old stellar population in the Galactic bulge.
With the aim of improving their near-infrared photometric characterization,
we investigate their near-infrared light curves, as well as the empirical relationships
between their light curve and metallicities using machine learning methods.

We introduce a new, robust method for the estimation of the
light-curve shapes, and hence the average magnitudes of RR~Lyrae variables in
the $K_\mathrm{S}$ band, by utilizing the first few principal components (PCs) as basis vectors,
obtained from the PC analysis of a training set of light curves. Furthermore,
we use the amplitudes of these PCs to predict the light-curve shape of each star
in the $J$-band, allowing us to precisely determine their average magnitudes (hence colors), even
in cases where only one $J$ measurement is available.

Finally, we demonstrate that the $K_\mathrm{S}$-band light-curve parameters of RR~Lyrae variables,
together with the period, allow the estimation of the metallicity of individual
stars with an accuracy of $\sim 0.2 - 0.25$\,dex, providing valuable chemical information about
old stellar populations bearing RR~Lyrae variables. The methods presented here can be straightforwardly
adopted for other classes of variable stars, bands, or for the estimation of other
physical quantities.

\end{abstract}




\section{Introduction} \label{sec:intro}

RR~Lyrae variables are tracers of the old stellar populations of galaxies, conveniently identifiable due to their characteristic light curves
and well-defined mean brightnesses.
The near-infrared (near-IR) light curves of RR~Lyrae stars are diminished in amplitude and
contain fewer features than at optical wavelengths,
making them challenging to discover. This is especially true of first-overtone (RRc) and double-mode (RRd) subtypes. Therefore,
in this study, we are only investigating the near-IR properties of fundamental-mode RR~Lyrae variables (RRab subtype, from here on RRLs).
Nevertheless, the lower absorption in the infrared bands, as well as the
slightly lowered effect of metallicity on the absolute magnitudes, makes
near-IR observations a tempting option for the determination of the
properties of old stellar populations. The reduced dependence on
metallicity in the near-IR is, however, still a subject of some debate. In
particular, the reduction is not seen in the calculations by
\citeauthor{2015ApJ...808...50M} (\citeyear{2015ApJ...808...50M}, see their Table 6),
whereas it is clearly present in the relations
provided by \cite{2004ApJS..154..633C}. Note that in the latter study,
evolutionary effects were fully taken into account, using synthetic horizontal branch
modeling; indeed, it is shown that one of the main advantages of
the near-IR regime is the reduced dependence on evolutionary effects, which
become increasingly more important as one moves toward bluer bandpasses.

Generally, to derive distances to these variables, their mean apparent magnitudes in at least two bands have to be determined,
allowing the estimation of the line-of-sight extinction toward each star. For classical radial pulsators, such as Cepheids and RRLs, it is customary
to fit the light curves with a truncated Fourier series (see, e.g., \citealt{1981ApJ...248..291S}), and to use the intercept of this fit as a measure
of the mean apparent magnitude.
The lower amplitudes and relatively high scatter of near-IR photometry present a challenge for this technique;
in time series with a limited amount of measurements, there are not enough data to accurately determine the coefficients of the high-order truncated
Fourier series needed to describe sharp features, such as the region of the minimum light of the light curves. Alternatively, light-curve templates,
such as those of \cite{1996PASP..108..877J}, could be used as model representations of the time series. However, this approach has drawbacks:
the \cite{1996PASP..108..877J} templates
only cover the $K$ band, and the $J$-band light curves of fundamental-mode RR~Lyrae are markedly different; with only four
RRL templates, not all possible light-curve shapes are represented. 

Although the effect of metallicity on the absolute magnitudes is lessened in the near-IR compared with the optical
(e.g., \citealt{2003MNRAS.344.1097B, 2004ApJS..154..633C}), knowledge of individual RRL
metallicities can still provide valuable insight into the star formation histories of the oldest populations of the Milky Way
in parts not accessible by optical spectroscopy and/or photometry. Relationships between
the light-curve shape, period, and the iron abundance [Fe/H], such as the widely used relation of \cite{1996AA...312..111J}, provide a convenient estimate
in the optical regime. Despite their obvious usefulness, no such relation has been established so far in the near-IR bands.

\setcounter{table}{0}
\begin{table*}[h!]
\renewcommand{\thetable}{\arabic{table}}
\centering
\caption{Collection of RR~Lyrae near-IR photometric observations} \label{tab:data}
\begin{tabular}{r@{\hskip4pt}l@{\hskip6pt}c@{\hskip6pt}c@{\hskip4pt}c@{\hskip4pt}l|r@{\hskip4pt}l@{\hskip6pt}c@{\hskip6pt}c@{\hskip4pt}c@{\hskip4pt}l|r@{\hskip4pt}l@{\hskip6pt}c@{\hskip4pt}c@{\hskip4pt}c@{\hskip4pt}l}
\tablewidth{0pt}
\hline
\hline
\multicolumn2c{ID$^{\mathrm{a}}$} & Period$^{\mathrm{b}}$ & J$^{\mathrm{c}}$ & Ref. & [Fe/H]$^{\mathrm{e}}$ &
\multicolumn2c{ID$^{\mathrm{a}}$} & Period$^{\mathrm{b}}$ & J$^{\mathrm{c}}$ & Ref. & [Fe/H]$^{\mathrm{e}}$ &
\multicolumn2c{ID$^{\mathrm{a}}$} & Period$^{\mathrm{b}}$ & J$^{\mathrm{c}}$ & Ref. & [Fe/H]$^{\mathrm{e}}$ \\
\hline
\decimals
            AV &   Peg                 &  0.390375   &  +  &  10      &$+0.08$ & $\omega$\,Cen  &  V055  &  0.581724   &  +  &  11  &...      & $\omega$\,Cen  &  V041  &  0.662942   &  +  &  11  &...  \\
          V445 &   Oph                 &  0.397020   &  +  &  4       &$+0.01$ & $\omega$\,Cen  &  V181  &  0.588370   &  +  &  11  &...      & $\omega$\,Cen  &  V013  &  0.669039   &  +  &  11  &...  \\
             W &   Crt                 &  0.412012   &  +  &  12      &$-0.45$ & $\omega$\,Cen  &  V025  &  0.588500   &  -  &  11  &...      & $\omega$\,Cen  &  V114  &  0.675307   &  +  &  11  &$-1.61$  \\
            AR &   Per                 &  0.425549   &  +  &  10      &$-0.14$ & $\omega$\,Cen  &  V045  &  0.589116   &  +  &  11  &...      & $\omega$\,Cen  &  V149  &  0.682728   &  +  &  11  &...  \\
            SW &   And                 &  0.442262   &  +  &  1,9,10  &$-0.06$ & $\omega$\,Cen  &  V125  &  0.592888   &  +  &  11  &$-1.81$  & $\omega$\,Cen  &  V046  &  0.686971   &  +  &  11  &...  \\
            RR &   Leo                 &  0.452390   &  +  &  10      &$-1.30$ & $\omega$\,Cen  &  V108  &  0.594458   &  -  &  11  &$-1.63$  & $\omega$\,Cen  &  V088  &  0.690211   &  +  &  11  &...  \\
$\omega$\,Cen  &  V112                 &  0.474359   &  +  &  11      &...     &             RV &   Phe  &  0.596400   &  +  &  2   &...      & $\omega$\,Cen  &  V102  &  0.691396   &  +  &  11  &$-1.65$  \\
            BB &   Pup                 &  0.480532   &  +  &  12      &$-0.35$ &             TT &   Lyn  &  0.597436   &  +  &  10  &$-1.50$  & $\omega$\,Cen  &  V097  &  0.691898   &  +  &  11  &$-1.74$  \\
$\omega$\,Cen  &  V130                 &  0.493250   &  -  &  11      &...     & $\omega$\,Cen  &  V033  &  0.602324   &  +  &  11  &$-1.58$  & $\omega$\,Cen  &  V007  &  0.713026   &  +  &  11  &...  \\
          WVSC &   054$^{\mathrm{d}}$  &  0.501267   &  +  &  5       &$-1.50$ & $\omega$\,Cen  &  V090  &  0.603404   &  +  &  11  &$-1.78$  &             VY &   Ser  &  0.714094   &  +  &  4,8 &... \\
$\omega$\,Cen  &  V074                 &  0.503209   &  +  &  11      &...     & $\omega$\,Cen  &  V049  &  0.604650   &  +  &  11  &...      & $\omega$\,Cen  &  V116  &  0.720074   &  -  &  11  &$-1.11$  \\
$\omega$\,Cen  & NV457                 &  0.508619   &  +  &  11      &...     &             UU &   Cet  &  0.606074   &  +  &  2   &...      & $\omega$\,Cen  &  V034  &  0.733967   &  +  &  11  &...  \\
$\omega$\,Cen  &  V023                 &  0.510870   &  +  &  11      &$-1.35$ & $\omega$\,Cen  &  V079  &  0.608276   &  +  &  11  &...      & $\omega$\,Cen  &  V172  &  0.738049   &  -  &  11  &...  \\
$\omega$\,Cen  &  V107                 &  0.514102   &  +  &  11      &...     & $\omega$\,Cen  &  V118  &  0.611618   &  +  &  11  &$-2.04$  & $\omega$\,Cen  &  V085  &  0.742758   &  +  &  11  &...  \\
          WVSC &   055$^{\mathrm{d}}$  &  0.514810   &  +  &  5       &$-1.50$ & $\omega$\,Cen  &  V020  &  0.615559   &  +  &  11  &$-1.52$  & $\omega$\,Cen  &  V109  &  0.744098   &  +  &  11  &$-1.70$  \\
$\omega$\,Cen  &  V005                 &  0.515274   &  -  &  11      &$-1.24$ & $\omega$\,Cen  &  V027  &  0.615680   &  +  &  11  &$-1.16$  & $\omega$\,Cen  &  V111  &  0.762905   &  +  &  11  &$-1.79$  \\
          WVSC &   047$^{\mathrm{d}}$  &  0.519611   &  +  &  5       &$-1.50$ & $\omega$\,Cen  &  V062  &  0.619770   &  +  &  11  &...      & $\omega$\,Cen  &  V099  &  0.766181   &  -  &  11  &$-1.91$  \\
$\omega$\,Cen  &  V008                 &  0.521329   &  +  &  11      &...     & $\omega$\,Cen  & NV458  &  0.620326   &  +  &  11  &...      & $\omega$\,Cen  &  V054  &  0.772915   &  +  &  11  &$-1.80$  \\
          WVSC &   046$^{\mathrm{d}}$  &  0.529820   &  +  &  5       &$-1.50$ & $\omega$\,Cen  &  V032  &  0.620347   &  -  &  11  &...      & $\omega$\,Cen  &  V038  &  0.779061   &  +  &  11  &$-1.64$  \\
$\omega$\,Cen  &  V120                 &  0.548537   &  +  &  11      &$-1.15$ & $\omega$\,Cen  &  V018  &  0.621689   &  +  &  11  &...      & $\omega$\,Cen  &  V026  &  0.784720   &  +  &  11  &$-1.81$  \\
          WVSC &   050$^{\mathrm{d}}$  &  0.551535   &  +  &  5       &$-1.50$ & $\omega$\,Cen  &  V096  &  0.624527   &  +  &  11  &...      & $\omega$\,Cen  &  V057  &  0.794402   &  +  &  11  &...  \\
$\omega$\,Cen  &  V100                 &  0.552745   &  +  &  11      &...     &             SS &   Leo  &  0.626344   &  +  &  4   &$-1.56$  & $\omega$\,Cen  &  V015  &  0.810642   &  +  &  11  &$-1.68$  \\
            RR &   Cet                 &  0.553038   &  +  &  10      &$-1.29$ & $\omega$\,Cen  &  V004  &  0.627320   &  +  &  11  &...      & $\omega$\,Cen  &  V268  &  0.812922   &  +  &  11  &$-1.76$  \\
            TU &   UMa                 &  0.557648   &  +  &  1       &$-1.15$ & $\omega$\,Cen  &  V115  &  0.630469   &  +  &  11  &$-1.64$  & $\omega$\,Cen  &  V063  &  0.825943   &  +  &  11  &...  \\
$\omega$\,Cen  &  V067                 &  0.564446   &  -  &  11      &$-1.19$ & $\omega$\,Cen  &  V146  &  0.633092   &  -  &  11  &...      & $\omega$\,Cen  &  V128  &  0.834988   &  +  &  11  &...  \\
$\omega$\,Cen  &  V044                 &  0.567545   &  +  &  11      &$-1.29$ & $\omega$\,Cen  &  V040  &  0.634072   &  +  &  11  &$-1.62$  & $\omega$\,Cen  &  V144  &  0.835320   &  +  &  11  &...  \\
$\omega$\,Cen  &  V056                 &  0.568023   &  -  &  11      &...     & $\omega$\,Cen  &  V122  &  0.634929   &  +  &  11  &$-1.79$  & $\omega$\,Cen  &  V003  &  0.841258   &  +  &  11  &...  \\
            SW &   Dra                 &  0.569670   &  +  &  7       &$-0.95$ &           NSV  &   660  &  0.636985   &  +  &  13  &$-1.31$  & $\omega$\,Cen  &  V411  &  0.844880   &  +  &  11  &...  \\
$\omega$\,Cen  &  V106                 &  0.569903   &  -  &  11      &$-1.90$ &              W &   Tuc  &  0.642230   &  +  &  2   &$-1.37$  & $\omega$\,Cen  &  V104  &  0.866308   &  +  &  11  &...  \\
            RV &   Oct                 &  0.571163   &  +  &  12      &$-1.08$ & $\omega$\,Cen  &  V086  &  0.647844   &  +  &  11  &$-1.99$  & $\omega$\,Cen  &  V091  &  0.895225   &  +  &  11  &$-1.81$  \\
$\omega$\,Cen  &  V113                 &  0.573375   &  +  &  11      &...     &              X &   Ari  &  0.651180   &  +  &  3,6 &$-2.10$  & $\omega$\,Cen  &  V150  &  0.899367   &  +  &  11  &...  \\
$\omega$\,Cen  &  V051                 &  0.574152   &  +  &  11      &$-1.84$ & $\omega$\,Cen  &  V134  &  0.652903   &  +  &  11  &...      & $\omega$\,Cen  & NV455  &  0.932517   &  +  &  11  &...  \\
            WY &   Ant                 &  0.574341   &  +  &  12      &$-1.39$ & $\omega$\,Cen  &  V069  &  0.653195   &  -  &  11  &...      & $\omega$\,Cen  &  V263  &  1.012158   &  +  &  11  &$-1.73$  \\
$\omega$\,Cen  &  V073                 &  0.575215   &  +  &  11      &...     & $\omega$\,Cen  &  V132  &  0.655656   &  -  &  11  &...      \\
\hline
\multicolumn{18}{p{1\textwidth}}{Notes.

$^{\mathrm{a}}$ Unique identifier of the variable.

$^{\mathrm{b}}$ Period in days.

$^{\mathrm{c}}$ Flag whether the $J$-band data are present and utilized in Section~\ref{subsec:j}.

$^{\mathrm{d}}$ The variables WVSC 054, 055, 047, 046 and 050 are V83, V116, V108, V55 and V40, respectively, from the globular cluster Messier 3
\citep{2001AJ....122.2587C, 2015AA...573A.100F}.

$^{\mathrm{e}}$ Sources of the iron abundances: $\omega$\,Cen stars: \cite{2006ApJ...640L..43S}; M3 stars: \cite{2009AA...508..695C}; NSV~660: \cite{2014ApJ...780...92S};
all other stars: \cite{1996AA...312..111J}.

References. (Photometric system): 1 -- \cite{1992PASP..104..514B} (CIT), 2 -- \cite{1992ApJ...396..219C} (ESO), 3 -- \cite{1989MNRAS.236..447F} (AAO),
4 -- \cite{1990MNRAS.247..287F} (SAAO), 5 -- \cite{2015AA...573A.100F} (WFCAM), 6 -- \cite{1987ApJ...312..254J} (CIT), 7 -- \cite{1987ApJ...314..605J} (CIT),
8 -- \cite{1988ApJ...332..206J} (CIT), 9 -- \cite{1992ApJ...386..646J} (CIT), 10 -- \cite{1989ApJS...69..593L} (CIT), 11 -- \cite{2015AA...577A..99N} (VISTA),
12 -- \cite{1993MNRAS.265..301S} (SAAO), 13 -- \cite{2014ApJ...780...92S} (2MASS). 
For the definitions of the photometric systems, we refer to \citeauthor{2018MNRAS.474.5459G} (\citeyear{2018MNRAS.474.5459G}, VISTA),
\citeauthor{2009MNRAS.394..675H} (\citeyear{2009MNRAS.394..675H}, WFCAM),
\cite{2001AJ....121.2851C} and references therein (all other systems). 
}
\end{tabular}

\end{table*}

The main motivation for the current study is the VISTA Variables in the V\'ia L\'actea (VVV) ESO Public Survey \citep{2010NewA...15..433M},
an extensive $K_\mathrm{S}$-band variability survey, also including $YZJH$ imaging,
conducted with the VIRCAM near-IR camera at the VISTA 4.1m telescope at Paranal Observatory,
Chile.
As it surveys the most crowded regions of the Milky Way, the Galactic bulge and southern disk, VVV has the capability of
uncovering previously uncharted parts of our Galaxy, utilizing different stellar tracers, such as Cepheids and RRLs.
For example, RRLs identified by the Optical Gravitational Lensing Experiment variability survey (OGLE; \citealt{2011AcA....61....1S})
have been combined with VVV photometry
to determine the distance of the Galactic bulge and to constrain the spatial distribution of its old component \citep{2013ApJ...776L..19D}.
Due to the limiting factor of the diminished amplitudes, searches for new RRLs in the VVV fields have been generally limited to
RRab variables.
Notwithstanding, directed searches have already led to the discovery of new RRLs \citep{2015A&A...575A.114G, 2016A&A...591A.145G, 2017AJ....153..179M}
in the VVV. These have also led to the discovery of a new Galactic globular cluster by the spatial clustering of RRLs
in the VVV disk fields \citep{2017ApJ...838L..14M}.
The need for the reproducible, automatic classification of RRLs has led to the development of a machine-learned classifier for finding RRab variables in the
$K_\mathrm{S} $-band \citep{2016A&A...595A..82E}, which enables the discovery of thousands of new RRLs
among the hundreds of millions of stellar sources of the VVV survey. 

In this article, we introduce several new methods, motivated by the desire of making maximum use of the RRLs in the VVV survey,
for the analysis of variable stars in general, and for the near-IR light curves of RRLs in particular.
Utilizing a high-quality RRL sample (Section~\ref{sec:train}), we apply principal component analysis (PCA) to the $K_\mathrm{S}$-band light curves
with the aim of decreasing the parameters required to accurately describe the various light-curve shapes of RRab variables (Section~\ref{sec:pca}).
We demonstrate how the $J$-band light-curve shape can be approximated, utilizing the $K_\mathrm{S}$-band principal component amplitudes, allowing the
determination of the $J$-band average magnitudes, even from a single observation (Section~\ref{subsec:j}).
Utilizing the first few principal components as basis vectors, we describe a robust nonlinear fitting technique (Section~\ref{sec:robust}).
Finally, we demonstrate, on a selected sample of OGLE-IV RRLs, that the light-curve shapes of RRLs in the $K_\mathrm{S}$ band,
together with
the pulsation period, can be used to estimate their metallicities, similar to their optical light curves (Section~\ref{sec:feh}).
The methods developed here are utilized in an accompanying paper \citep{dekany} for the characterization of the RRL population of the VVV disk fields.

\section{Model representation of near-IR RRab light curves} \label{sec:pca}

We have elected to utilize PCA to give a compact representation of the near-IR RRL light curves.
PCA is a widely used dimensionality reduction procedure to transform an original set of variables by an orthogonal transformation into
a new set of linearly uncorrelated variables called principal components (PCs). Generally, the first few PCs contain most of the variation
in the original data.
The procedure was first described more than a century ago by \cite{pearson1901}, and then rediscovered and named by \cite{hotelling1933}.
PCA has two main uses for data analysis: (1) reducing the number of dimensions of a data set by keeping only the most significant PCs and (2) identifying hidden
trends in the data. As astronomical data sets are inherently multidimensional (e.g., images, spectra, individual element abundances of stars, etc.), 
PCA has been adopted for both purposes by the astronomical community. A non-exhaustive list of PCA applications includes spectral classification of galaxies
\citep{1998A&A...332..459G}, stars \citep{1998MNRAS.295..312S}, and quasars \citep{2004AJ....128.2603Y}; modeling systematics in light curves \citep{2013ApJ...778..184J};
removal of galaxies from images with the aim of finding gravitionally lensed background galaxies \citep{2016A&A...592A..75P}; analysis of galaxy velocity curves
\citep{2017MNRAS.469.2539K}; as well as looking for correlations between diffuse interstellar band features \citep{2017ApJ...836..162E}.

In the context of variable stars, the most important applications of PCA for the study of Cepheids and RR~Lyrae stars were
done by \cite{2002MNRAS.329..126K}, \cite{2004MNRAS.355.1361K}, \cite{2005MNRAS.363..749T}, and \cite{2017MNRAS.466.2805B}, while \cite{2009A&A...507.1729D} analyzed a range
of variable star classes with the aim of evaluating PCs as a metric for light curves in large databases.

\subsection{The light-curve training set} \label{sec:train}

We have collected high-quality $K_\mathrm{S}$ and $J$-band photometry available from the literature for 101 RRLs.
Table~\ref{tab:data} summarizes the training set.
The available data can be categorized into three main types:
near-IR photometry taken with the aim of Baade-Wesselink analysis of RRLs \citep{1992PASP..104..514B, 1992ApJ...396..219C,
1989MNRAS.236..447F, 1990MNRAS.247..287F, 1987ApJ...312..254J, 1987ApJ...314..605J, 1988ApJ...332..206J, 1992ApJ...386..646J, 1989ApJS...69..593L,
1993MNRAS.265..301S};
serendipitous observations of RRLs in the 2MASS \citep{2014ApJ...780...92S} and WFCAM \citep{2015AA...573A.100F} calibration fields;
and the extensive $J$ and $K_\mathrm{S}$ variability study of $\omega$\,Centauri by \cite{2015AA...577A..99N}.
There are other sources of near-IR time-series photometry available in the literature for RRLs
(see, e.g., \citealt{2014A&A...567A.100A}), but we decided to utilize only photometry
where the phase coverage and quality of the observations are adequate for the accurate determination
of the near-IR light-curve shapes, at least in the $K_\mathrm{S}$ band.

Examining Table~\ref{tab:data} reveals that the variables in the field of $\omega$\,Centauri \citep{2015AA...577A..99N} dominate the sample.
Although this could introduce a heavy bias toward a particular metallicity, $\omega$\,Cen contains multiple stellar populations
(e.g., \citealt{2011A&A...533A.120V, 2012A&ARv..20...50G}; and references therein),
with two contributing to the RRL sample (\citealp{2006ApJ...640L..43S}, and references therein).
These two populations have [Fe/H] $\sim -1.2$ and $\sim -1.7$.
The variables utilized from the photometry of \cite{2015AA...573A.100F} are all members of the globular cluster M3, with a metallicity
of $\sim -1.5$ \citep{2009AA...508..695C}. Furthermore, the field RRLs have a wider metallicity distribution,
allowing us to cover the physical parameter space of RRLs.

As for the period distribution, the sample covers most of the possible period range of RRLs, from 0.39 to 1.01 days. While very metal
rich RRLs can have periods as short as about 0.35 days, their optical light-curve shapes are not drastically different from those with periods around
0.4 days. Therefore, we can surmise that the present data set can be considered representative of RRLs and their light-curve shapes.

\subsection{Data preparation} \label{subsec:dataprep}

\begin{figure*}[]
\epsscale{1.15}
\plotone{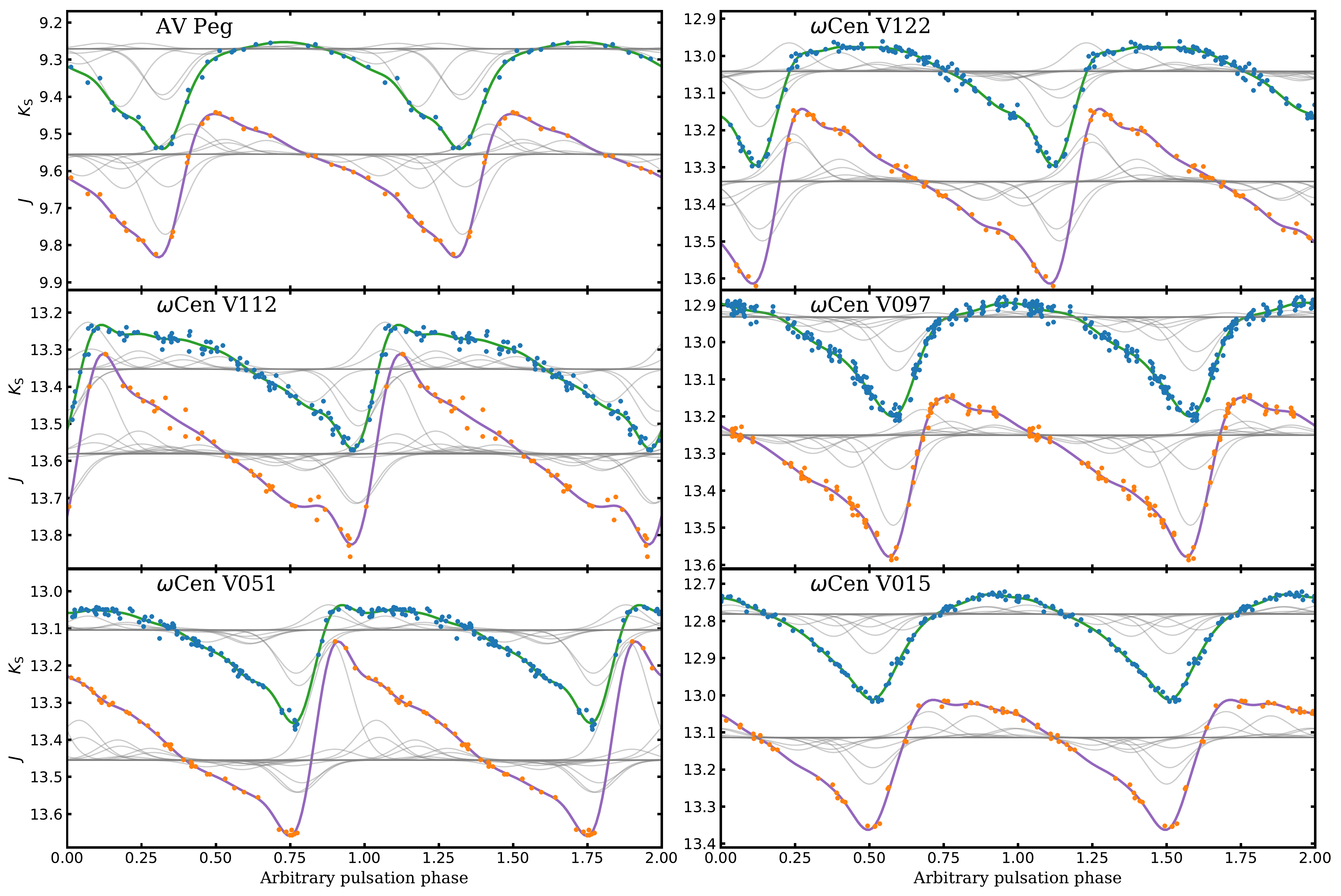}
\caption{Typical folded $K_\mathrm{S}$ (top, blue points) and $J$-band (bottom, orange points) light curves of our RRL sample.
Each light curve is modeled with a sum of periodic Gaussian (von Mises) basis functions, following Eq.~\ref{sum}.
As we utilize LASSO regularization, most periodic Gaussians have zero amplitudes.
The individual periodic Gaussians with non-zero amplitudes are illustrated by the faint
gray lines for each band and variable, while the green and purple curves illustrate the light curve fit (sum of the
individual periodic Gaussians) in the $K_\mathrm{S}$ and $J$-band, respectively.}
\label{fig:lasso}
\end{figure*}

To apply PCA to our data outlined in Section~\ref{sec:train}, we have to describe their phase-folded light-curve shape. As mentioned in
Section~\ref{sec:intro}, it can be hard to accurately represent the near-IR RRL light curves as a Fourier series.
Furthermore, our light curves have vastly different numbers of data points: NSV~660 has almost 3000, while for the RRL found in $\omega$~Centauri,
\cite{2015AA...577A..99N} only acquired a total of 100 and 42 epochs in the $K_\mathrm{S}$ and $J$ bands, respectively.

As an alternative, folded light curves can be described as the linear sum of a series of different basis functions of choice, such as Gaussians, which can be
aligned to the phased light-curve points with ordinary least squares (OLS) regression. We have chosen to adopt the Gaussian sum fitting example
of Fig.~8.4 of \cite{2014sdmm.book.....I}, to the $K_\mathrm{S}$ and $J$-band RRL light curves.
As RRL light curves are strictly periodic, we have decided to replace the Gaussians with their periodic analog, the circular normal
(also called Von Mises) distributions of the form

\begin{equation}
f(x) = \frac{e^{\kappa \cos(x-\mu)} }{2\pi I_0 (\kappa)}, 
\end{equation}

\noindent where $\mu$ is the measure of location (analogous to the mean of the Gaussian distribution), $\kappa$ is the measure of concentration (where $1/\kappa$ is
analogous to the variance, $\sigma^2$ of a Gaussian), and $I_0(\kappa)$ is the modified Bessel function of order 0. To model the light-curve shapes,
we define the sum of 100 of these basis functions, distributed evenly between phases 0 and 1:

\begin{equation}
\label{sum}
\mathrm{LC} = m + \sum_{i=0}^{99} A_i \frac{e^{\kappa \cos[2\pi (x-\frac{i}{100})]} }{2\pi I_0 (\kappa)},
\end{equation}

\noindent where $A_i$ are the individual amplitudes of the circular normal distributions, and $m$ is the intercept of the fit.
Although we could use OLS to find the amplitudes of Eq.~\ref{sum}, most of the light curves have less than 100 points
available, leading to an underdetermined problem. In such cases, regularization can be introduced to penalize the
magnitude of independent parameters (in our case, the amplitudes $A_i$), by modifying the loss function.
We do so by utilizing the least absolute shrinkage and selection operator (LASSO or L1 regularization; \citealt{LASSO}),
which adds the sum of absolute values of the regression coefficients multiplied by a regularization parameter $\alpha$
to the loss function. We note that by utilizing LASSO, generally most coefficients end up being zero\footnote{Another popular
choice for these kinds of problems is the Tikhonov regularization, also known as L2 regularization
or Ridge regression, where the sum of the squares of the fit coefficients times $\alpha$ is added to the loss function. In
contrast to LASSO, Ridge regression does not result in sparse solutions (i.e., most parameters do not end up being zero),
making the interpretation of the results of the fit harder to interpret.}.

We have utilized the linear regression routines of \textsf{scikit-learn} \citep{scikit-learn} to fit the folded $K_S$ and $J$
light curves of each RRL with Eq.~\ref{sum} utilizing LASSO regularization. Our fit has two hyperparameters:
$\kappa$ and $\alpha$. We have utilized both cross validation (leave-one-out for stars with few light-curve points,
N-fold otherwise) and manual inspection of the resulting light-curve fits to determine the optimal values of these
parameters. High values for the concentration parameter $\kappa$ grants our model the ability to fit sharp features,
such as the rising branches of certain variables, but can cause an overfit in phase ranges with few points. Conversely,
at low values of $\kappa$ the model cannot fit sharp features. We have found that a numerical value of $\kappa=6$ is optimal for our model
for both the $K_\mathrm{S}$ and $J$-band light curves of RRLs. As for the regularization parameter $\alpha$, our
cross validation resulted in different optimal values for different stars, typically in the range between $10^{-4}$ and $10^{-5}$.
As visual inspection did not reveal significant differences when changing the regularization parameter between these two
values, we have adopted an intermediate value of $10^{-4.5}$ for all of our stars.

\begin{figure}[]
\epsscale{1.1}
\plotone{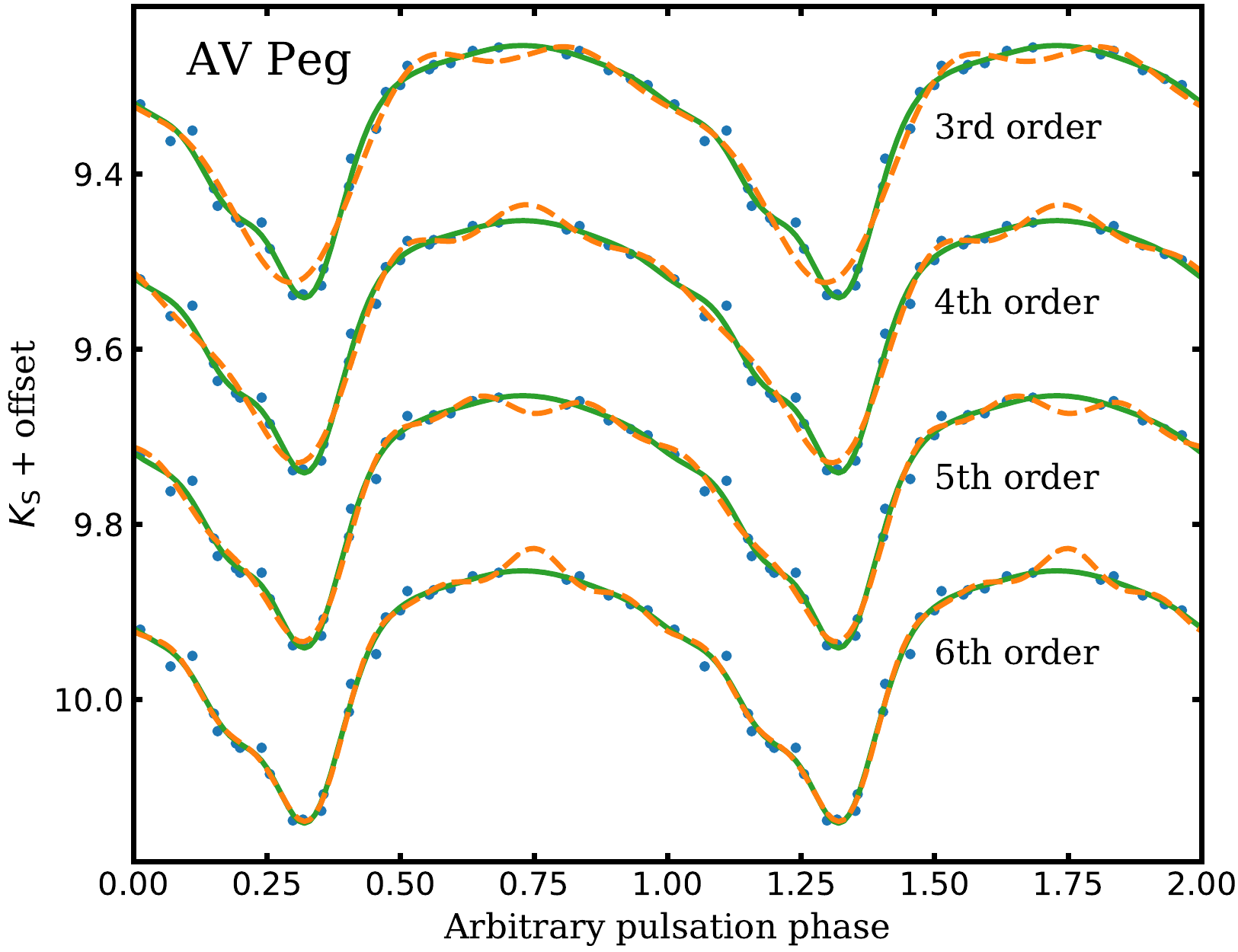}
\caption{Comparison between our fit with the circular normal distributions (Eq.~\ref{sum}, continuous lines), and Fourier series of different orders
(dashed lines).}
\label{fig:fourier}
\end{figure}

Figure~\ref{fig:lasso} illustrates the quality of the light curves and their fits. During our fitting process, some outlying points
have been removed manually and the periods of some variables have been revised, when tension existed between different photometric
sources. Table~\ref{tab:data} contains the revised periods for all variables. 
Figure~\ref{fig:fourier} compares our fit with Fourier series of different orders. As can be seen, our method provides a better
representation for variables with light-curve gaps.

The utilized light curves had been obtained in a variety of photometric systems, as detailed in Table~\ref{tab:data}.
Therefore, the resulting light-curve fits were all transformed to the photometric system of VISTA.
For variables not in the VISTA, WFCAM, or 2MASS systems, first they were transformed
to the system of 2MASS utilizing the updated transformation
formulae\footnote{\url{http://www.astro.caltech.edu/~jmc/2mass/v3/transformations/}}
of \cite{2001AJ....121.2851C}. Then, the 2MASS and WFCAM photometry was transformed to the VISTA system with the help of
the CASUVERS~1.4 transformations given by the Cambridge Astronomical Survey
Unit (CASU)\footnote{\url{http://casu.ast.cam.ac.uk/surveys-projects/vista/technical/photometric-properties}
The transformations between the VISTA and WFCAM systems were carried out using the relations updated on 2014 July 30.
Very recently, updated transformations were provided by \cite{2018MNRAS.474.5459G}. We have checked that the changes in
the resulting magnitudes are less than 0.005~mag. In addition, because this only affects the five stars from
\cite{2015AA...573A.100F}, none of the results in this paper are significantly affected by the choice of transformation
equations between these two systems. }.

\begin{figure}[]
\epsscale{1.1}
\plotone{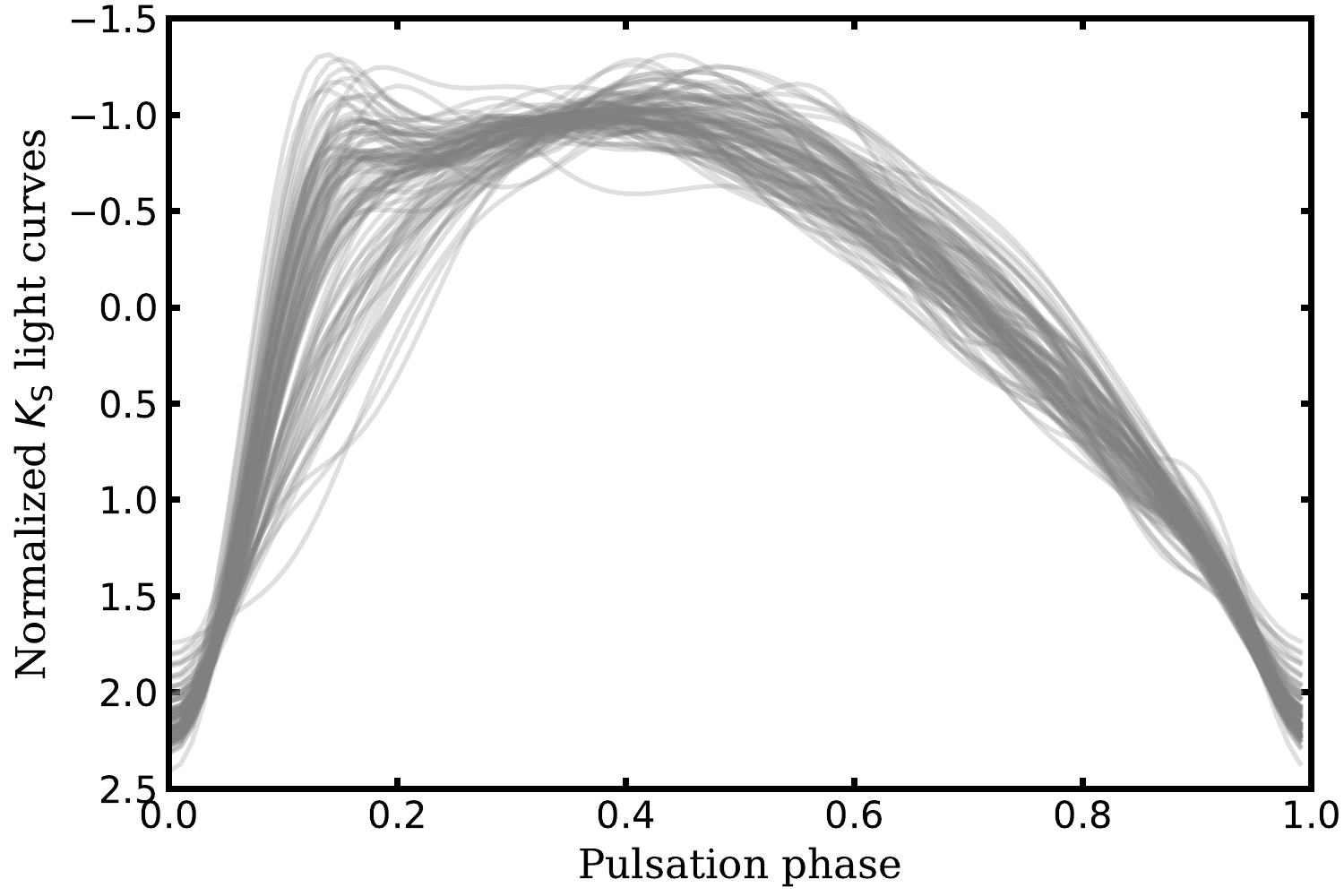}
\caption{Normalized, folded, and minima-aligned $K_S$-band light curves of our RRL sample.}
\label{fig:k}
\end{figure}

\subsection{Application of PCA on the $K_\mathrm{S}$-band light curves} \label{subsec:pca_app}

To apply PCA, we sample our light-curve fits on a grid of 100 phase points, evenly distributed
phases from 0.0 to 0.99. In the analysis of pulsating variables, it is customary to set the light-curve maxima at phase 0.0;
however, inspecting the $K_\mathrm{S}$-band light curve examples of Figure~\ref{fig:lasso} reveals that the maxima of RRLs
in the $K_\mathrm{S}$ band are ill defined: the timing of the maxima of the light curve depends heavily on the strength of
the bump on the rising branch. Therefore, we have chosen to align our light curves by the much sharper feature
of the light-curve minima (similar to the case of eclipsing binaries). Furthermore, in PCA, the sample values (the magnitudes in our case)
are usually
normalized to a mean of 0 and scatter of 1 along each dimension (phase). As our goal is to describe the light-curve shapes of
RRLs in the $K_\mathrm{S}$ band as a linear combination of PCs, we have chosen to normalize each light curve independently
to have a mean of 0 and scatter of 1. These aligned, normalized input light curves can be seen in Figure~\ref{fig:k}.

\begin{figure}[]
\epsscale{1.1}
\plotone{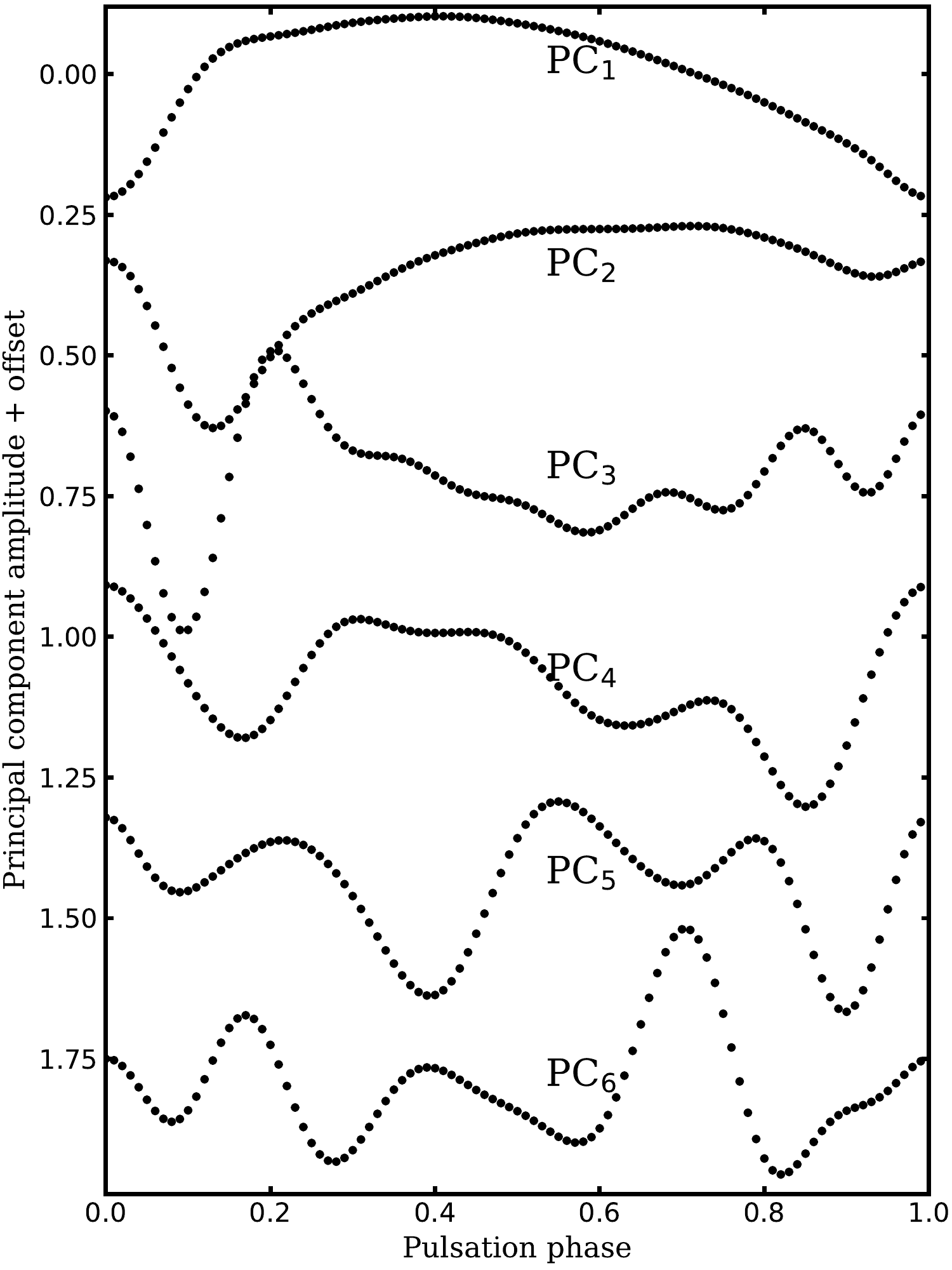}
\caption{From top to bottom: PCs 1 through 6, according to our decomposition of the $K_\mathrm{S}$-band RRL light curves.}
\label{fig:pca_comps}
\end{figure}

We carry out PCA by utilizing singular value decomposition, adopted from the PCA module of \textsf{scikit-learn} \citep{scikit-learn}.
Figure~\ref{fig:pca_comps} shows the first six PCs, according to our decomposition. As we have chosen not to normalize in each phase point,
the first PC contains the average light-curve shape of the normalized light curves. Including further components to describe the
light curves modifies this average shape, and this can be easily understood in the context of the individual light curves,
for PCs of low order: the second component can make the bump at the end of the rising branch (around phase 0.15) more or less
pronounced, while the third component is important to reproduce the double-peaked light curves displayed by some of the variables.

The power of the PCA lies in the fact that generally the first few PCs, together with their amplitudes, are sufficient to describe
the original input data, and the rest of the PCs can be discarded. The number of significant PCs can be decided by examining the
fraction of the variance explained by the components. As is obvious, the first PC dominates the
explained variance, because we normalized each light curve individually instead of normalizing the magnitudes along each dimension (phase).
If we look at the variance explained by the rest of the PCs, they explain 3.68, 1.39, 0.96, 0.72, etc. percent of the total variance, or
88.8, 3.4, 2.3, 1.7, etc. percent of the residual variance, when the variance explained by the first PC is subtracted from the total variance.
On the basis of these values, we deem that the first four PCs are sufficient for describing the $K_\mathrm{S}$-band light-curve shapes
of RRab stars.

\begin{figure}[]
\epsscale{1.1}
\plotone{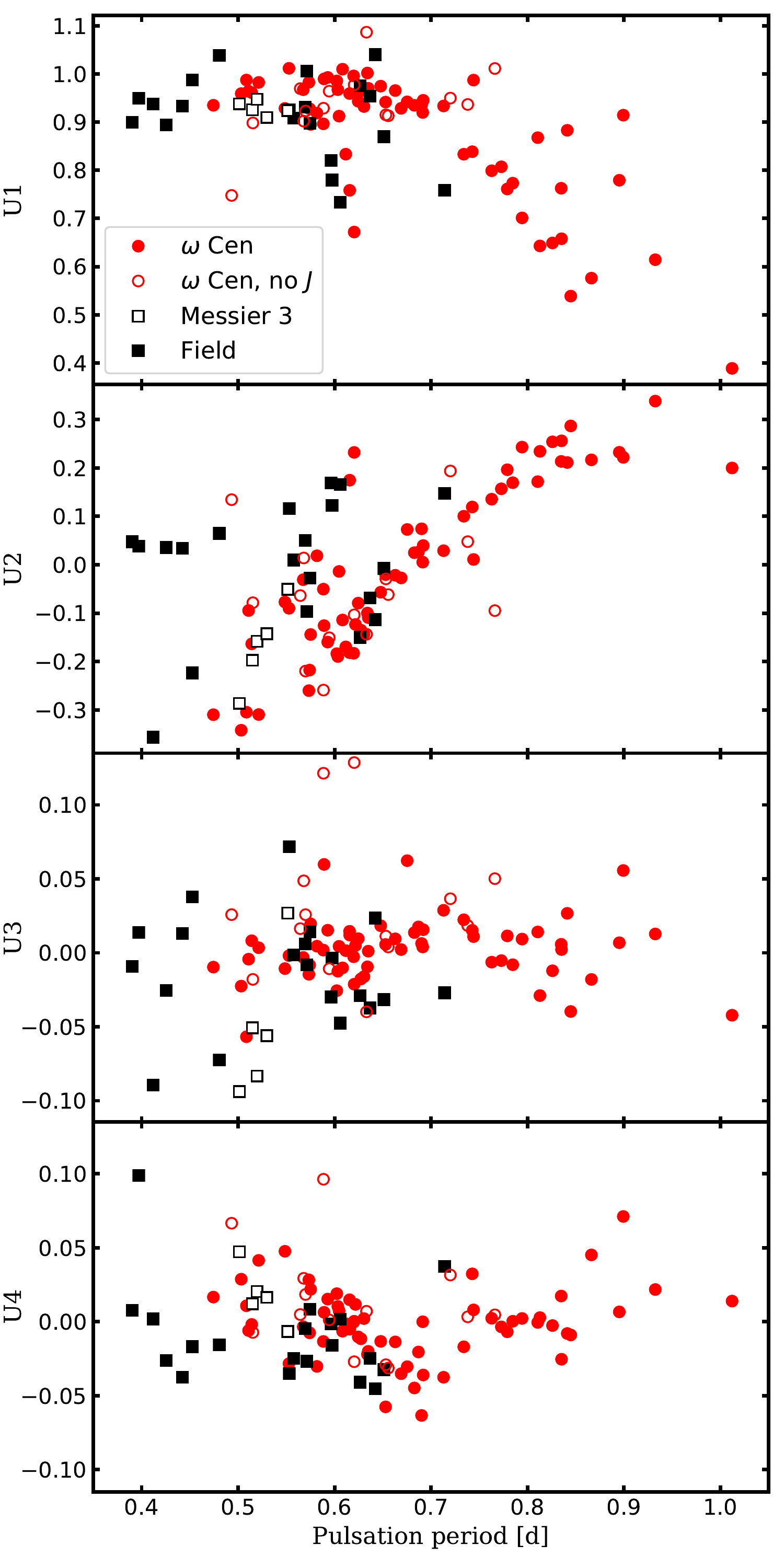}
\caption{Amplitudes $U_i$ of the PCs for the training sample. Circles and squares denote RRLs from the $\omega$~Centauri
photometry of \cite{2015AA...577A..99N} and other sources, respectively. Empty squares denote RRLs in Messier 3 from the photometry of
\cite{2015AA...573A.100F}, while empty circles denote variables not considered for the approximation of the $J$-band light-curve shapes
in Section~\ref{subsec:j}.}
\label{fig:pca_amps}
\end{figure}

We applied the PCA method on the normalized light curves of our sample of variables to emphasize the light-curve shape differences
among RRLs in the $K_\mathrm{S}$ band, instead of the different pulsation amplitudes of each variable.
Consequently, the individual amplitudes (also called eigenvalues) of the first four PCs,
$u_{1j}$, $u_{2j}$, $u_{3j}$ and $u_{4j}$, where $j=1..101$ is the index of the variables, carry no amplitude information on the
original light curves. By multiplying these amplitudes with the normalization constants used to normalize each of the light curves, or by
utilizing OLS to directly align the PCs to the light curves, we can determine the PC amplitudes
$U_{1j}$, $U_{2j}$, $U_{3j}$ and $U_{4j}$ for each star in our sample.
The sum of the PCs multiplied with these amplitudes recover the original light-curve shapes (and amplitudes) with high accuracy.

The distribution of the amplitudes $U_{ij}$ is illustrated in Figure~\ref{fig:pca_amps}.
The shapes of the light curves of RRLs, represented by the amplitudes of these first few PCs, potentially contain
information on the physical properties of the variables themselves, when the pulsation periods of the stars are taken into account.
The first amplitude, $U_1$ (from now on, we omit
the $j$ index for simplicity), can be viewed as
an analog to the total amplitude in amplitude-period (also called Bailey) diagrams. As described before, additional PCs modify
the light-curve shape. Keeping this in mind, it is immediately obvious that, at a given period, the amplitudes $U_1$ and $U_2$ separate
two sequences, which we can associate with the Oosterhoff type I and II groups of variables
(Oo; \citealt{1939Obs....62..104O}; see \citealt{2015pust.book.....C}, for a recent review and references).
As these groups at least partly
correlate with metallicity, we are going to explore the
possibility of estimating the metallicity of individual RRLs when their $K_\mathrm{S}$-band light curves are described by
the PC amplitudes $U_{i}$, in Section~\ref{sec:feh}.

\subsection{Approximation of the $J$-band light-curve shape}\label{subsec:j}

\begin{figure}[]
\epsscale{1.1}
\plotone{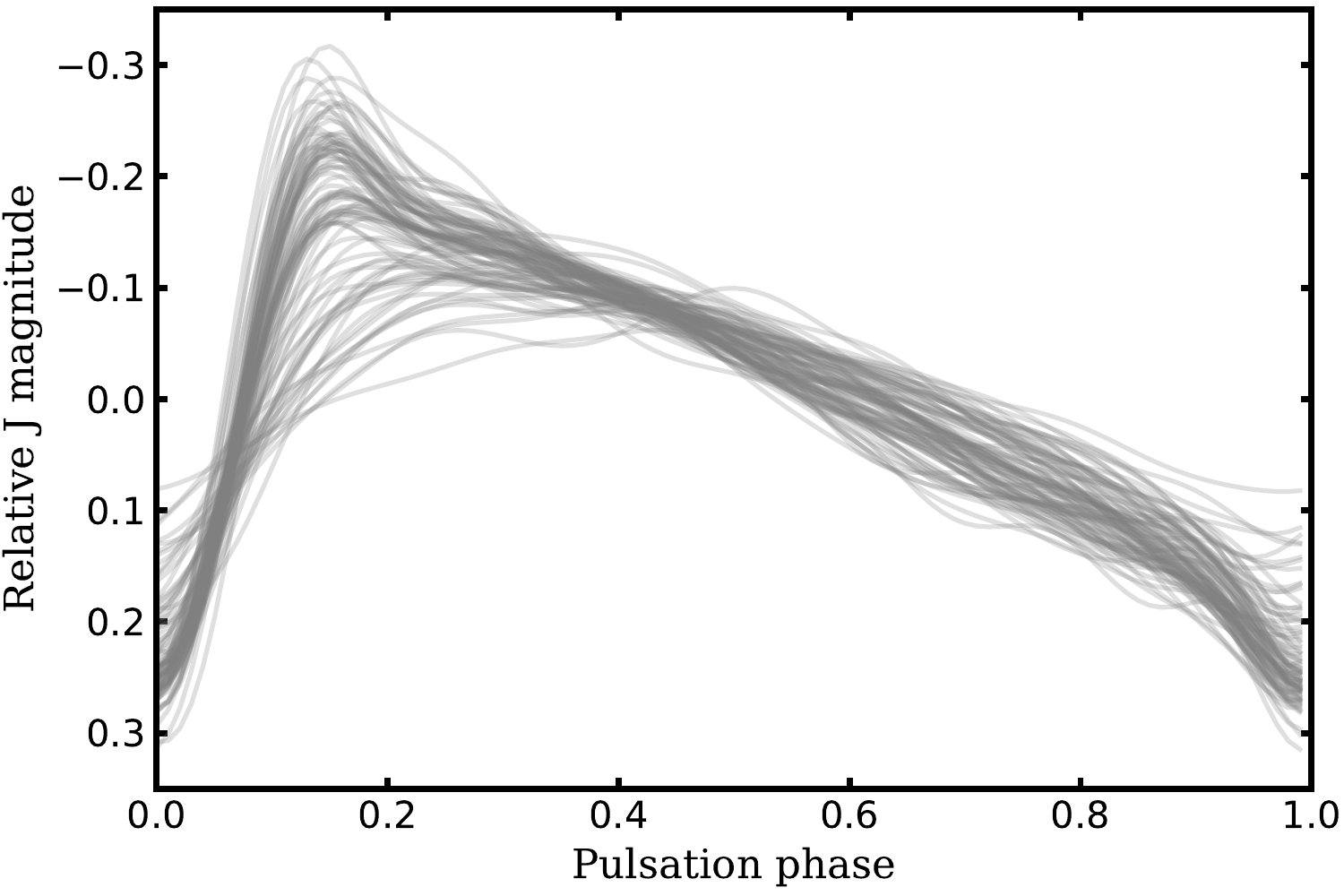}
\caption{Phase- and average-aligned $J$-band light curves of the training sample utilized for the approximation of the $J$-band light-curve shapes.
Note that these light curves are aligned in phase by the $K_\mathrm{S}$-band light-curve minima.
Only data for stars with high-quality $J$-band light curves, marked with a `+' sign in column 3 of Table~\ref{tab:data}, are shown.}
\label{fig:j}
\end{figure}

The light-curve shapes and amplitudes of pulsating variables vary among bandpasses, due to the difference in the relative contribution of the
change of radius and photospheric temperature during the pulsation cycle to the emitted flux at different wavelengths
(\citealt{2015pust.book.....C}, and references therein). Comparing the $K_\mathrm{S}$
and $J$-band light curves in Figure~\ref{fig:lasso} clearly demonstrates this.

During the course of the VVV survey, each field has been
observed only a few times in the $J$-band.
To determine accurate mean $J$ magnitudes for the RRLs, it is necessary to
describe the light-curve shapes of the stars in the $J$-band, i.e., the deviation of the $J$-band magnitude from its average in each
phase of the pulsation cycle. Usually, the difference between the light-curve shapes is ignored, and the $K_\mathrm{S}$-band light-curve shape
is used to estimate the average $J$-band magnitude. However, this method, depending on the light-curve phases of the observations, introduces
additional scatter in the derived magnitudes.

As the PC amplitudes $U_i$ provide a concise description of the $K_\mathrm{S}$-band light-curve shapes, we can evaluate whether the $J$-band
light-curve shapes can be approximated with their use. Besides these amplitudes, we consider the period as a possible additional parameter, to
assess its effect on variables that otherwise possess the same light-curve shapes in the $K_\mathrm{S}$ band.


The $\omega$~Centauri photometry of \cite{2015AA...577A..99N} contains only 42 epochs in the $J$-band; therefore, depending on their periods,
some variables have gaps in their $J$-band light curves at critical phases. A few other stars possess photometric anomalies in their light curves,
preventing us from obtaining a good light-curve fit for them in this band. We decided to omit these stars, marked in in Table~\ref{tab:data}, and
continue with the analysis of the remaining 87 RRLs.
These $J$ light curves
are first sampled in the same 100 phase points, as has been done for the $K_\mathrm{S}$-band light curves in Section~\ref{subsec:pca_app}, where
phase 0.0 is the phase of the $K_\mathrm{S}$-band light-curve minimum for each of the variables\footnote{We could reach an incorrect solution
if the photometry presented in Table~\ref{tab:data} would have $K_\mathrm{S}$ and $J$-band photometries combined from different epochs,
as an incorrect period or (slow) period change could affect the phases. However, as all stars have simultaneous photometry in these two bands, this is not a problem
in our case.}.
These curves are aligned to have
a mean of 0, but in contrast to the PCA analysis of the $K_\mathrm{S}$-band light curves, they are not normalized.

The training set of $J$-band light curves is shown in Figure~\ref{fig:j}. We are searching for the ideal set of variables
to describe these light-curve shapes, e.g., the deviations from the mean in each phase. We can consider this as a separate
problem at each light-curve phase for which we are looking for a linear model description, as a combination of yet to be determined parameters as the basis. The connection
between the $J$ magnitudes in different light-curve phases is that these linear models should depend on the same parameters.
Our candidate parameters are the period $P$, the four PC amplitudes of the $K_\mathrm{S}$-band light curves $U_1$, $U_2$, $U_3$ and $U_4$,
as well as their polynomial combinations up to the third order ($P^2$, $PU_1$, $PU_2$, $PU_3$, $PU_4$, $U_1^2$, etc.).

\begin{figure}[]
\epsscale{1.1}
\plotone{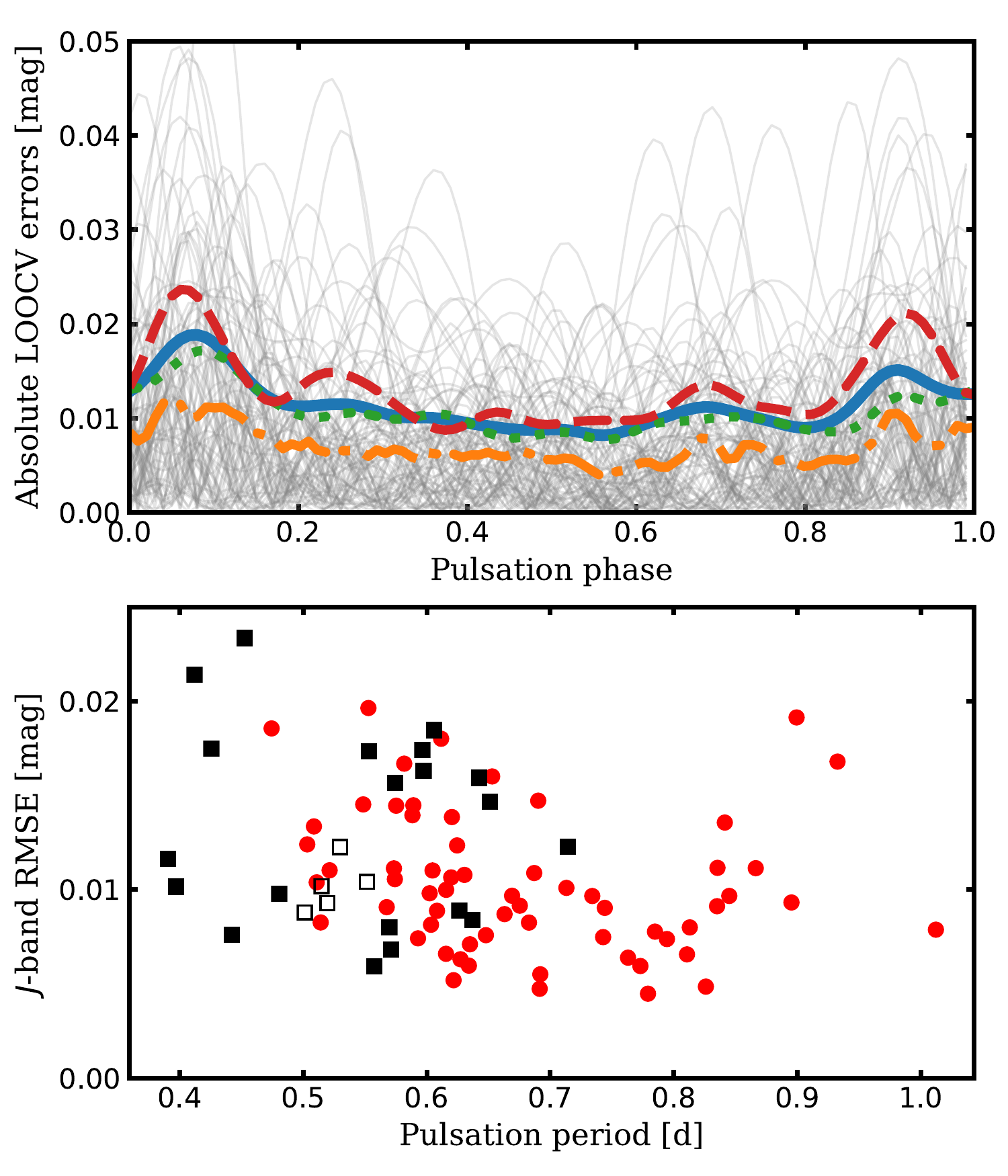}
\caption{Top: absolute $J$-band residuals for our approximation of the $J$-band light-curve shape using the PC amplitudes,
as a function of the pulsation phase.
Thin gray lines show the individual absolute residuals, according to our LOOCV. The continuous, dashed, and dotted
lines mark the average errors for the complete sample, the stars of $\omega$ Centauri, and the sample omitting the
$\omega$ Centauri stars, respectively. The dotted-dashed line shows the median error for each phase bin.
Bottom: $J$-band RMSEs from our LOOCV. The symbols are the same as in Fig.~\ref{fig:pca_amps}.
}
\label{fig:residuals}
\end{figure}

We performed an exhaustive search for the best parameter combination, where we considered various permutations containing up to six candidate parameters.
As our sample size is fairly small, we have chosen not to utilize a hold-out set, nor N-Fold Cross Validation for the validation of our results,
as the exact (random) choice of the hold-out set or the N-Folds could lead to heavily biased results (for example, if all the short period variables
are selected to be in one fold).
Therefore, for each candidate parameter set, we have opted to perform Leave-One-Out Cross Validation (LOOCV) the following way:

\begin{enumerate}
\item we evaluate 87 separate sub-cases, where we omit the light curve of each of the 87 variables;
\item in each case, we optimize a linear solution using OLS for the $J$-band magnitudes, using the candidate parameter set as basis;
\item with these solutions, we approximate the $J$-band light curve of the omitted variable, and calculate the residuals; and
\item for each candidate parameters set, we assess the total root mean square error (RMSE) as a performance metric.
\end{enumerate}

\begin{figure*}[]
\epsscale{1.1}
\plotone{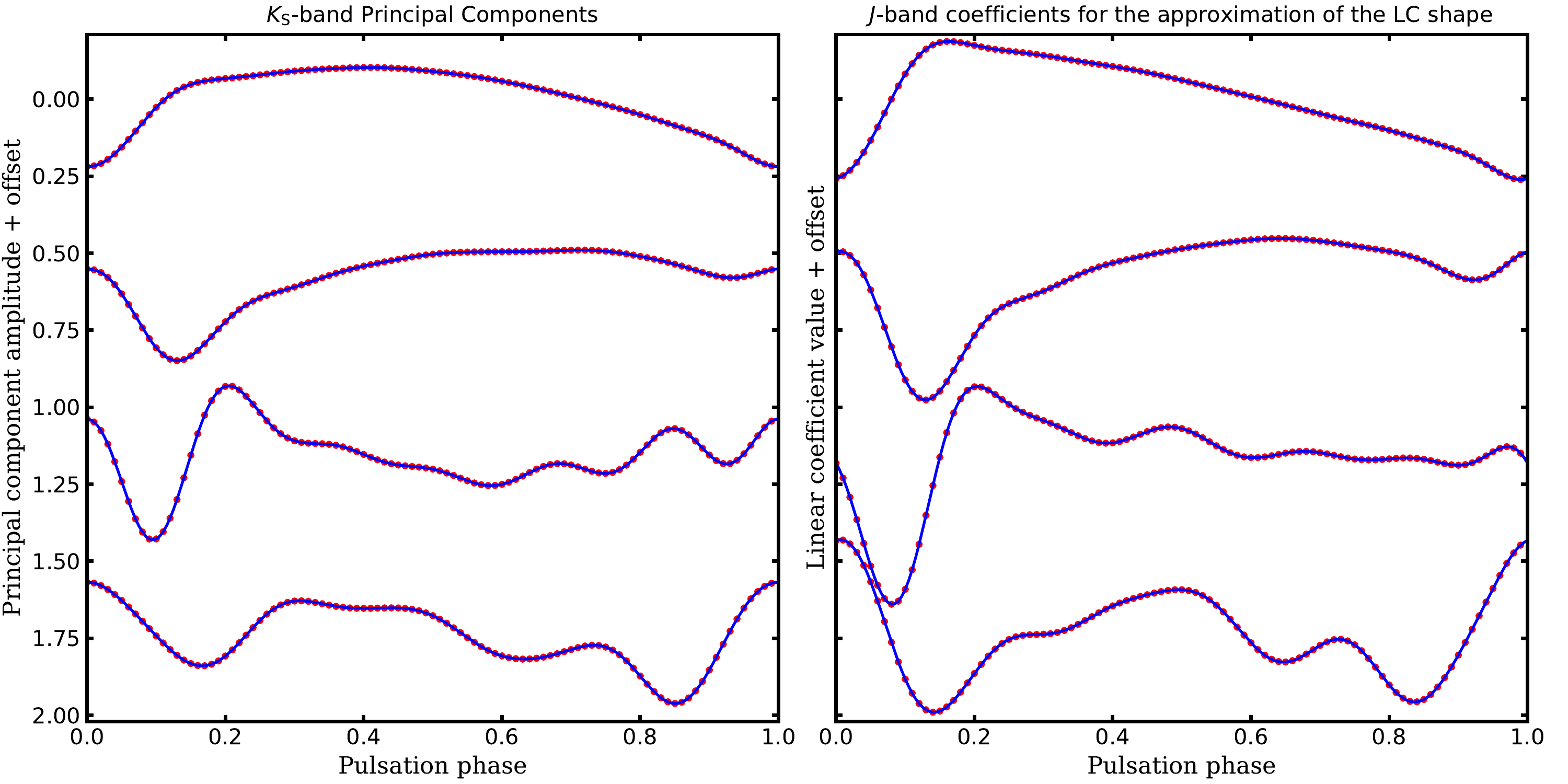}
\caption{Left: The first four principal components (red circles) and their continuous Fourier representations
(blue lines), utilizing Eq.~\ref{pca_fourier}.
Right: the coefficients for the approximation of the $J$-band light-curve shape in each phase (red points)
and their Fourier representations (blue lines). Note that the coefficients utilized to reconstruct the $J$-band
light curves describe similar light-curve features as the $K_\mathrm{S}$-band PCs, albeit with larger amplitudes,
especially around the rising branch. This behavior is in line with the fact that RRLs have larger amplitudes at
shorter wavelengths.}
\label{fig:pca_harm_j_pred}
\end{figure*}

We have found that out of all the possible parameter combinations, by only considering the PC amplitudes of the $K_\mathrm{S}$-band light curves
$U_1$, $U_2$, $U_3$ and $U_4$, we reach a total RMSE of our linear prediction of just 0.012\,mag. 
Although this value can 
still be lowered by including combinations with the period, such as $PU_1$, we noticed that this mainly decreases the errors for variables coming
from $\omega$ Centauri, where multiple stars have very similar periods and $K_\mathrm{S}$ and $J$ light-curve shapes, while the errors for
other RRLs, especially the short period ones, increase drastically.
Figure~\ref{fig:residuals} illustrates the residuals both as a function of the pulsation phase (top panel) and for individual stars from the LOOCV
(bottom panel).

Based on our analysis, we conclude that the $J$-band light-curve shapes of RRLs can be estimated using their $K_\mathrm{S}$-band PCA
amplitudes with high precision. Figure~\ref{fig:pca_harm_j_pred} illustrates this process by comparing the $K_\mathrm{S}$-band PCs derived
in Section~\ref{subsec:pca_app} (left-hand panel) with the coefficients found by OLS for the prediction of the $J$-band light-curve shape
(right-hand panel),
when all of the 87 RRLs with good $J$ light curves are considered.

\section{Robust fitting of RRL $K_\mathrm{S}$-band light curves} \label{sec:robust}

Our goal is to provide an accurate and convenient method for determining the light-curve shapes, periods and
average magnitudes of RRLs in the VVV survey.
As we have seen in Section~\ref{sec:pca}, the PC amplitudes $U_i$ provide a compact description
of the $K_\mathrm{S}$-band light-curve shape; furthermore, the $J$-band light-curve shapes can also be
approximated using the same $U_i$ coefficients as obtained from the $K_\mathrm{S}$-band data.

The PCs themselves have been utilized before by \cite{2005MNRAS.363..749T} in the case of Cepheids,
to produce a series of realistic template light curves in the optical, which could be used to determine
specific parameters, such as the periods or the magnitude
at maximum light, from relatively scarce photometry.
In contrast to the study of \cite{2005MNRAS.363..749T}, we implement a method to fit the PCs directly to the VVV $K_\mathrm{S}$-band light curves.
In Section~\ref{subsec:pca_app}, we determined the PCs on a grid of light-curve phases. However, in order to fit a target light curve,
we have to discern their PC amplitudes $U_i$, using the PCs as basis functions. As PCs are not continuous functions, we transform them
to a Fourier representation of the form

\begin{equation}
\label{pca_fourier}
\mathcal{F}_{PC, i} (x) = \sum_{k=1}^{12} \left[A_{ik} \sin(2\pi kx)     +  B_{ik} \sin(2\pi kx) \right],
\end{equation}

\noindent where $x$ is the light-curve phase, using OLS fitting of the individual PCs.
The continuous lines on the left panel of Fig.~\ref{fig:pca_harm_j_pred} demonstrate the transformed PCs.
These new representations can be used as basis functions to describe the light-curve shapes
of RRLs as

\begin{equation}
\label{lc_model}
\mathrm{LC}(t) = m_0 + \sum_{i=1}^{4} \left[ U_i \times \mathcal{F}_{PC, i} \left(\frac{t-E_0}{P}\right) \right],
\end{equation}

\noindent where $t$ is the Julian Date, and the free parameters are the PC amplitudes $U_i$, the magnitude average $m_0$,
the period $P$ and the zero epoch $E_0$.

The light-curve representation proposed by Eq.~\ref{lc_model} is not a linear function; therefore, we have to use nonlinear
regression.
We developed a method to adjust these seven free parameters, and the model presented in Equation~\ref{lc_model}, to the light curves of
RRLs in the VVV survey. Unfortunately, in the more crowded fields and near bright stars, VVV photometry tends to contain a
fraction of outlying points. Traditionally, iterative threshold rejection is utilized to omit outlying measurements in massive time-series
analysis. However, in cases of unfortunate data distribution, and due to the fact that in the first iteration erroneous observations can have a
large effect on the fit, this can result in the flagging of good data as outliers.
To avoid this problem, we replace the normally used squared error loss with the Huber loss function \citep{huber} of the form

\begin{equation}
\label{huber}
L_\delta(y,f(x)) = 
    \begin{cases}
        \frac{1}{2}\big(y-f(x)\big)^2 & \quad \mathrm{for}\, | y-f(x) < \delta | \\
        \delta | y-f(x) | -\frac{1}{2} \delta^2 & \quad \mathrm{otherwise}.
    \end{cases}
\end{equation}

\noindent This loss function behaves identically to the squared error loss for data points with residuals smaller than $\delta$.
For residuals bigger than $\delta$, the loss grows linearly with increasing residuals. Therefore, outlying points weigh less than
they would, if squared error loss was utilized, a convenient feature in the case of outliers.

\begin{figure*}
\epsscale{1.12}
\plotone{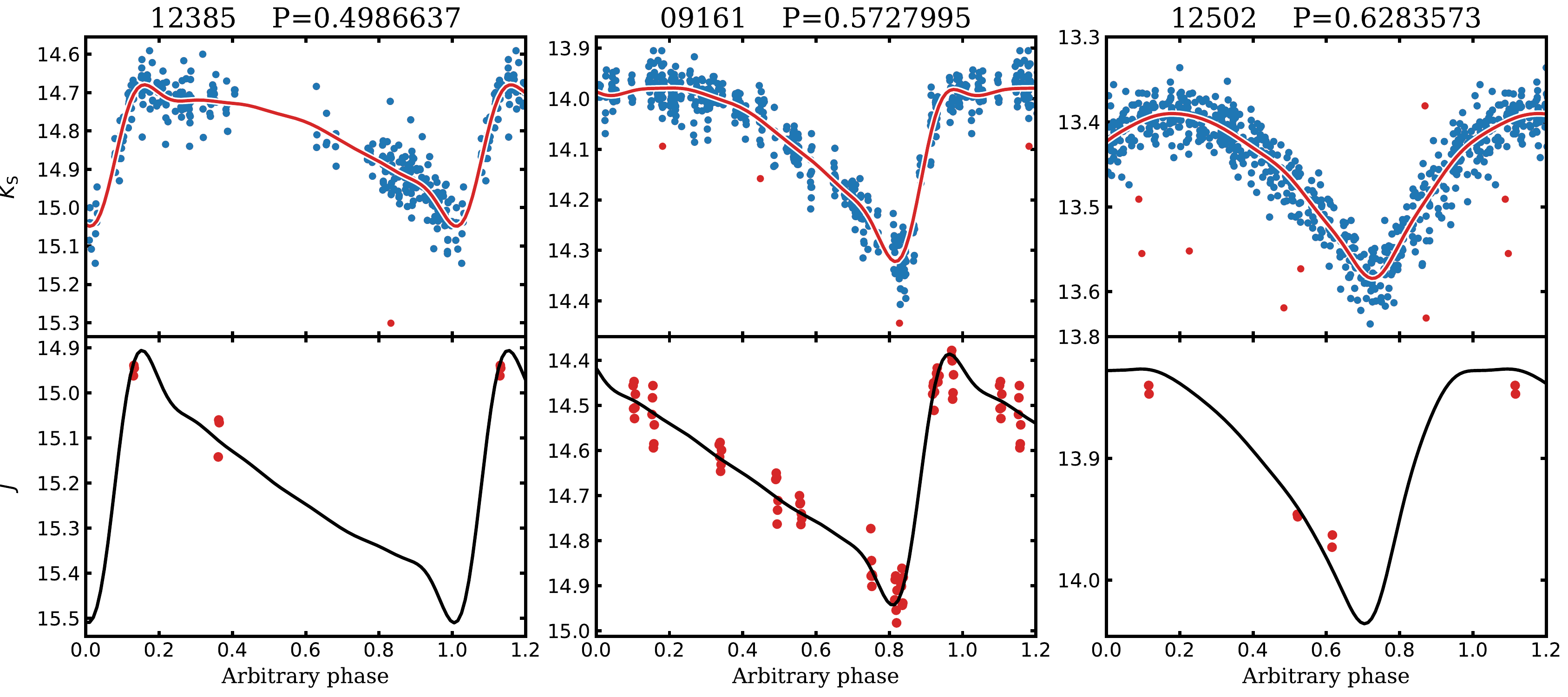}
\caption{Examples of the near-IR light-curve fitting methods implemented in Sect.~\ref{sec:robust}. OGLE IDs and periods are shown on the top of the
panels.
Top: $K_\mathrm{S}$-band light curves and their fits using the first four PCs derived in Sect.~\ref{sec:pca}.
Bottom: $J$-band light-curve points and the approximations of the $J$-band light-curve shapes from the PC amplitudes, as described in Sect.~\ref{subsec:j}.
Variable no.~12385 (left panel) illustrates the case of a single large gap in the folded light curve, a challenge for traditional Fourier fitting.
No.~09161 has an unusually large number of $J$-band light-curve points, allowing us to demonstrate the quality of the predicted $J$ light curve.
No.~12502 has a much more symmetric light-curve shape, but our method is flexible enough to find a good fit for it.
}
\label{fig:examples}
\end{figure*}

The distribution of PC amplitudes of our PCA sample (Fig.~\ref{fig:pca_amps}) gives us \textit{a priori} information on the possible
shapes of RRL light curves, if we accept the sample analyzed in Sect.~\ref{sec:pca} as representative of RRLs.
We utilize this information as follows. We require the first PC amplitude to be in the range
$U_1 \in [U_{1,\mathrm{max}} + \delta_{U_1}, U_{1,\mathrm{min}}- \delta_{U_1}]$, where $U_{1,\mathrm{max}}$ and $U_{1,\mathrm{min}}$
are the largest and smallest $U_{1}$ amplitudes of the PCA training sample (see Sect.~\ref{subsec:pca_app}), respectively, and
$\delta_{U_1} = (U_{1,\mathrm{max}}-U_{1,\mathrm{min}})/10$.
In other words, we limit the $U_1$ amplitudes to not be larger or smaller than the largest or smallest $U_1$ amplitude in the PCA sample,
$\pm10\%$ of the total range of $U_1$ amplitudes given by the PCA. As for the other amplitudes, we require their relative values,
$U_2/U_1$, $U_3/U_1$ and $U_4/U_1$, to adhere to analogous requirements on their ranges. These requirements ensure that the fit
light curves have shapes that are similar to those in the PCA sample.

The following steps have been implemented in order to ensure that the light-curve shapes and $JHK_\mathrm{S}$ average magnitudes are derived with the
highest possible accuracy, based on the VVV data:

\begin{enumerate}

\item all $K_\mathrm{S}$-band light-curve points brighter than 9\,mag, fainter than 20\,mag, as well as 
points farther than $\pm0.5$\,mag from the median are discarded;

\item the remaining light-curve points are fit with the light-curve model of Eq.~\ref{lc_model}, while utilizing the Huber loss function
in the form of Eq.~\ref{huber}, with $\delta=0.05$\,mag, and the constraints on the ranges of $U_1$ and $U_{2,3,4}/U_1$, as described above; \label{enu:fit} 

\item we omit points with a residual larger than $3.5\sigma$, where $\sigma$ is determined from the absolute median deviation of the fit
from step~\ref{enu:fit};

\item the fit is repeated on the remaining data in the same way as described in step~\ref{enu:fit};

\item as the light-curve shapes of RRLs are very similar in the $H$ and $K_\mathrm{S}$ bands
(e.g., they are the same with a scatter of $\sim0.02$\,mag, depending on the pulsation phase;
see Figs.~1 and 2 of \citealt{1992PASP..104..514B}), we use the $K_\mathrm{S}$
light-curve shapes to determine the average magnitudes by fitting them directly to the $H$-band measurements; and

\item the $J$-band light-curve shapes are predicted using the $U_i$ magnitudes, as described in Sect.~\ref{subsec:j}, and these are fit to the $J$-band
measurements to determine the mean $J$ magnitude.

\end{enumerate}

The principal outputs of this procedure are the $K_\mathrm{S}$-band light-curve parameters, as well as the robust magnitude estimates
of the target RRLs in the $JHK_\mathrm{S}$ bands. Figure~\ref{fig:examples} illustrates the VVV $K_\mathrm{S}$-band fits (top),
as well as the predicted $J$-band light curves of three RRLs from the OGLE bulge RRL sample of \cite{2014AcA....64..177S}.
The implementation of all these steps in a single, convenient routine
is available on \textsf{GitHub}\footnote{\url{https://github.com/gerhajdu/pyfiner}}.

\section{Metallicity estimation from $K_\mathrm{S}$-band photometry}\label{sec:feh}

The idea of estimating the metallicity of RRLs from their Fourier light-curve parameters
originates from \cite{1995A&A...293L..57K}. Their original method was substantially improved by
\cite{1996AA...312..111J}, who found a simple linear relationship between the
iron abundances of RRLs, their periods and the epoch-independent Fourier phase differences
$\phi_{31}$ ($=\phi_3 - 3\phi_1$, where $\phi_3$ and $\phi_1$ are the Fourier phases of the
third and first harmonics of the light curve, respectively; \citealt{1981ApJ...248..291S})
of their $V$-band light curves.
Their formula was calibrated by spectroscopic measure ments and
corresponds to a metallicity scale established by high-dispersion spectroscopy \citep{1995AcA....45..653J} .
Following the example set by \cite{1996AA...312..111J}, \cite{2005AcA....55...59S}
developed similar formulae for the Cousins $I$-band, notable as the band in which
most observations of the OGLE surveys are being carried out \citep{2015AcA....65....1U}.
\cite{2005AcA....55...59S} gave two alternative formulae for the derivation of the
iron abundance, a 2- and a 3-term formula, and he argued that
the latter, which also includes the amplitude of the second Fourier harmonic,
provided better results. We emphasize that both the $V$- and $I$-band formulae
have a residual scatter of only $\sim$0.14~dex, which is similar to the accuracy of
spectrophotometric methods, such as the $\Delta S$ method
\citep{1959ApJ...130..507P}, and are on the same metallicity scale defined by
\cite{1995AcA....45..653J}.

A similar calibration between [Fe/H] and the near-IR light-curve parameters has
long been lacking, despite the fact that with the advent of large time-domain
near-IR photometric surveys, such as the VVV and the VISTA Survey of the Magellanic System
\citep{2011A&A...527A.116C}, among others,
such an empirical calibration is of key
importance, as a large fraction of the newly discovered distant RRL stars along
the Galactic plane are beyond the faint magnitude limit of optical surveys.
In the following, we are going to detail our calibration of such a relationship,
utilizing the overlapping RRLs found by the OGLE project \citep{2014AcA....64..177S}
and observed by the VVV survey.

\subsection{VVV photometry of bulge RR~Lyrae variables} \label{subsec:fits}

We employed our light-curve fitting method described in Section~\ref{sec:robust} to determine the
$K_\mathrm{S}$-band light-curve parameters of RRab variables listed in the OGLE-IV bulge RR~Lyrae catalog
\citep{2014AcA....64..177S}. Toward this end, we made use of the data processed 
by the VISTA Data Flow System (VDFS; \citealt{2004SPIE.5493..401E}), provided by CASU.
The aperture photometry of detector frame
stacks (pawprints)
was extracted using the Starlink Tables Infrastructure Library (STIL; \citealt{stil}) at the coordinates
of the OGLE RRLs.
This process has resulted in near-IR light curves for 24217 RRLs.

The VDFS provides photometry in circular apertures of different radii, but due to the source crowding in the
Galactic bulge fields, the adopted radii have to be chosen carefully.
 Generally, smaller apertures provide better photometry for dimmer objects, but in
relatively uncrowded cases, this does not always hold true. Furthermore, variables found in overlapping
regions of tiles and pawprints can have different offsets between them, but generally the best aperture
also minimizes this offset. Therefore, for each star, the optimal aperture has to be chosen on
a case-by-case basis.

We utilized our fitting procedure described in Sect.~\ref{sec:robust} on the five smallest
apertures provided by the VDFS for each OGLE RRL. It has been found that the final value of the
Huber cost function divided by the number of data points after the 3.5$\sigma$ clip
is a good indicator of the quality of the photometry obtained with a given aperture with respect to the others.
Hence, in this section we are going to utilize
the PC amplitudes determined for each star using the aperture for which this value is the smallest.

\subsection{Unbiased photometric metallicities from the RR~Lyrae $I$-band light curves} \label{subsec:met}

Due to the large number of RRLs with both OGLE-IV and VVV photometry,
we decided to search for correlations between the abundances determined on the former, and the light-curve
shapes of the latter.
As the main bulge population of RRLs have spectroscopic iron abundances of [Fe/H]$\sim -1$
\citep{1991ApJ...378..119W}, this component is expected to be dominant in the calculated
photometric metallicities as well.

The OGLE-IV (or OGLE-III, if the former was not available) $I$-band light curves of RRLs
in the common OGLE/VVV sample
were fit with a sixth-order Fourier series to calculate the Fourier parameters necessary for the
metal abundance formulae of \cite{2005AcA....55...59S}.

As our goal is to provide a method for estimating metallicity based on the near-IR light-curve parameters, we
eliminate all variables where either estimate is uncertain.

By far, the most common reason for the elimination of variables was the presence of the Blazhko effect,
which leads to distorted light-curve shapes in RRLs, which are never equivalent to those of non-modulated RRLs \citep{2002A&A...390..133J}.
\cite{2017MNRAS.466.2602P} studied the Blazhko effect on a subsample of bulge RRLs in the OGLE sample, and
found an incidence rate of $40\%$. We have inspected each $I$-band light curve and its fit visually to reveal the
presence of the Blazhko effect. In dubious
cases, we have inspected the Discrete Fourier Transform of the residual light curve for signs of the characteristic
side peaks of modulation in the Fourier domain. We opted to eliminate all variables from our selection where
inspection of the light curve gave hints of the Blazhko effect, which accounted for approximately the same
fraction of stars as found by \cite{2017MNRAS.466.2602P}. Furthermore, this inspection
revealed some dubious RRL light curves, as well as many light curves where the Fourier parameters cannot be
determined with high accuracy (faint variables, too few light-curve points, gaps in the folded light curve, etc.),
which were also removed in order to preserve the purity of the sample.

Besides requiring good-quality light curves in the $I$-band, the $K_\mathrm{S}$-band data were also subjected to
the following quality cuts: we discarded all variables where either the first PC amplitude, $U_1$,
or the PC amplitude ratios $U_2/U_1$, $U_3/U_1$, or $U_4/U_1$ exceeded the limit described in Sect.~\ref{sec:robust},
as these indicate problematic photometry in the VVV data, such as severe blending.

\begin{figure}[]
\epsscale{1.1}
\plotone{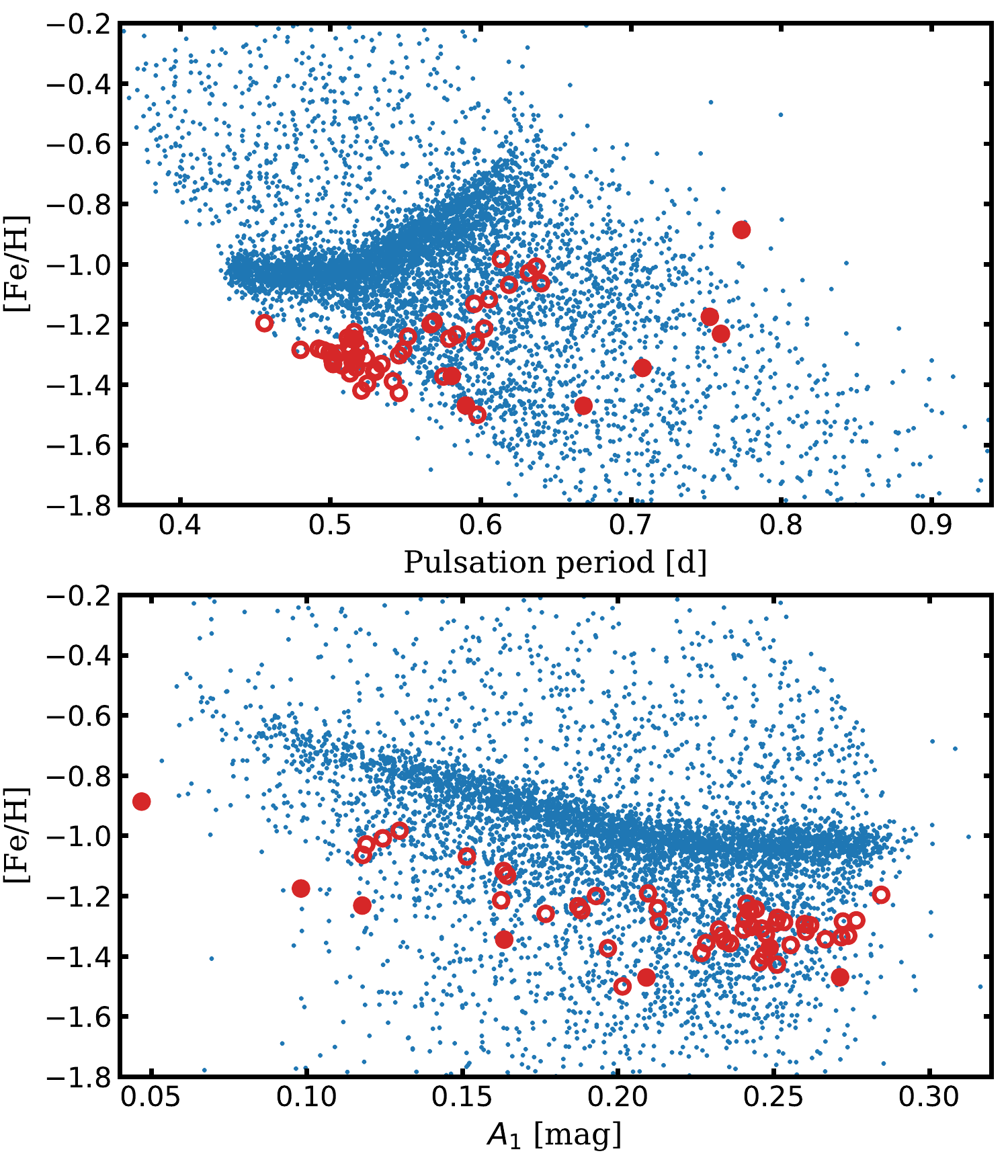}
\caption{ Photometric iron abundances calculated from the $I$-band light-curve parameters with Eq.~2 of \cite{2005AcA....55...59S}
as a function of the pulsation period (top) and as a function of the amplitude of the first harmonic of the
Fourier series fit (bottom), both times indicated by small blue points. Oo~I and II variables from M3 are overplotted
as large empty and filled circles, respectively. Both the bulge field and the M3 RRLs
display a systematic bias toward higher abundances as the pulsation period increases and the pulsation amplitude decreases.
Furthermore, at a given period
or amplitude, the metallicities derived for the Oo~I and II stars in M3 (calculated from the photometry of
\citealt{2017MNRAS.468.1317J})
at similar amplitudes are systematically offset.
}
\label{fig:feh}
\end{figure}

We utilized Eqs.~2~and~3 of \cite{2005AcA....55...59S} to calculate the metal abundances of the 6215 remaining variables.
The top panel of Figure~\ref{fig:feh} reveals a striking systematic on top of the overall linear trend in the photometric
metallicity calculated with Eq.~2 of \cite{2005AcA....55...59S}
as a function of the pulsation period. In the main locus of stars, denoting the Oosterhoff I population of the bulge RRLs,
longer-period RRLs seem to have systematically higher metallicities than their shorter-period counterparts.
Moreover, a similar trend can be seen on the bottom panel: the lower-amplitude stars have systematically higher metallicities.

\begin{figure}[]
\epsscale{1.1}
\plotone{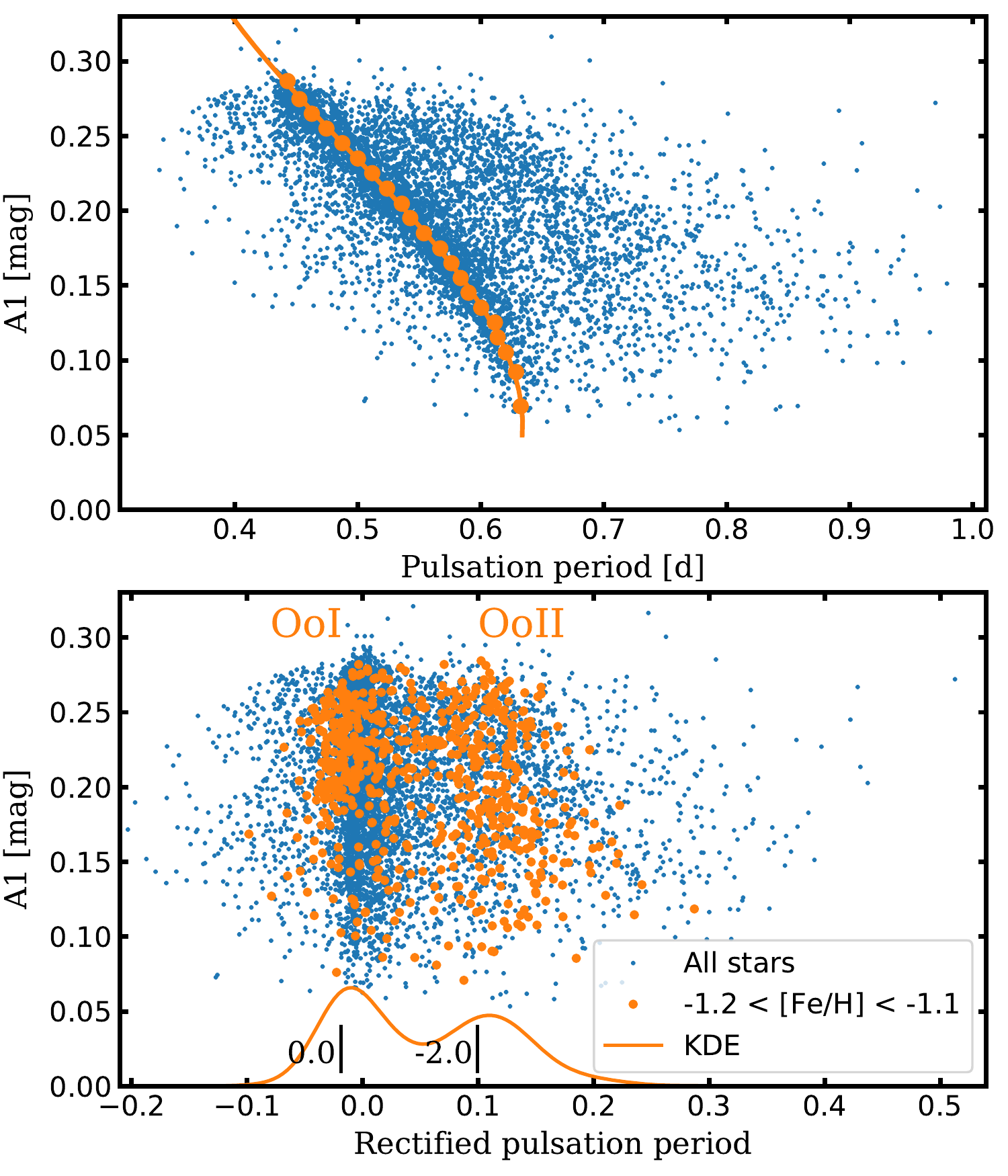}
\caption{ Iron abundance-dependent separation of Oosterhoff classes.
Top: the distribution of the amplitudes of the first harmonic of the $I$-band Fourier fits of bulge RRLs as a function
of the pulsation periods. The main ridge of Oo~I stars is localized using the kernel density estimate (KDE) maxima in different
amplitude bins (circles). These are fit with a third-order polynomial, illustrated by the continuous line.
Bottom: same as above, but as a function of the rectified pulsation periods, calculated as the difference between the
real pulsation period and the position of the Oo~I ridge of the top panel, as marked by the continuous line. The RRLs in the
abundance bin $-1.2<\textrm{[Fe/H]}<-1.1$ are marked, and the continuous line at the bottom shows their KDE distribution.
The local minima in the middle of the KDEs of different abundance bins change, reflecting the metallicity dependence of
the Oosterhoff phenomenon. We classify RRLs as Oo~I or II based on their position in this diagram, using a criterion that
is a linear function of their calculated iron abundance. The decision criteria for $\textrm{[Fe/H]}=0.0$ and $-2.0$ are marked
by the vertical black lines.
}
\label{fig:oo}
\end{figure}

\begin{figure}[]
\epsscale{1.1}
\plotone{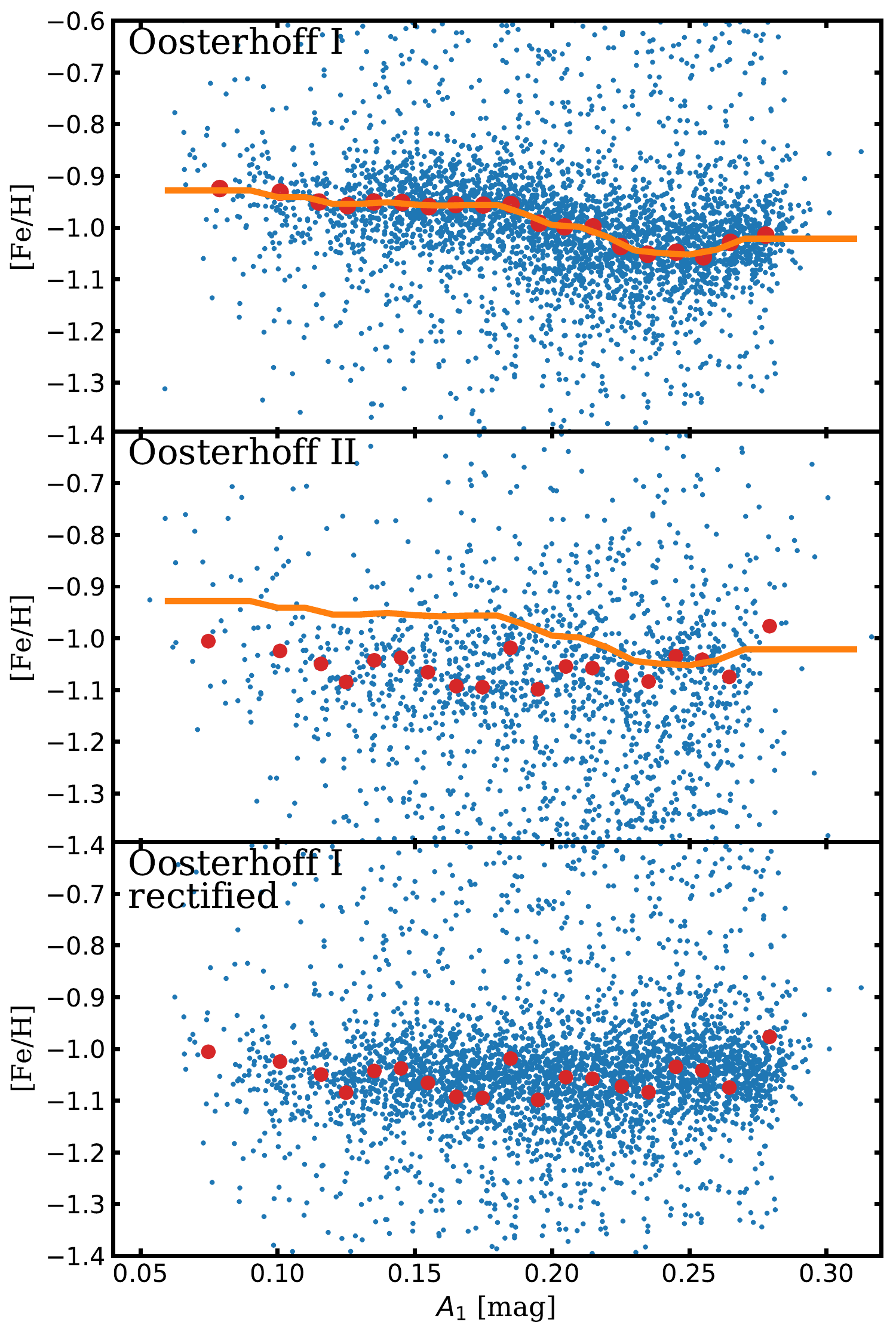}
\caption{ Photometric iron abundances (small blue points) calculated from the $I$-band light-curve parameters with
Eq.~3 of \cite{2005AcA....55...59S}.
Top: the photometric metallicity estimates of Oosterhoff I variables. As can be seen, these still display a dependency
on the amplitude of the first Fourier harmonic, albeit to a much smaller degree than found using Eq.~2 of \cite{2005AcA....55...59S}
(Fig.~\ref{fig:feh}). Red filled circles denote the average metallicities, as calculated from the maximum value of a KDE
of data points in different amplitude bins.
$K$-nearest regression (with $K=2$, marked as a continuous orange line) was utilized to generalize
this effect, and this function was used to rectify the individual abundance estimates.
Middle: photometric metallicity estimates of Oosterhoff II variables. The KDE estimates
(red filled circles) of the amplitude bins reveal no
significant dependency on the estimated metallicities. The contiguous orange line from the top panel is drawn for
comparison purposes.
Bottom: the rectified metallicity estimates of the Oosterhoff I RRLs. The KDE estimates from the middle panel
(filled red circles) are repeated for comparison purposes.
}
\label{fig:feh2}
\end{figure}

In globular clusters, for both Oosterhoff groups of fundamental-mode variables, the amplitude decreases with
increasing periods.
It is hard to conjure up a scenario where systematically higher-metallicity RRLs would only populate the
lower-amplitude, longer-period part of the main locus on this
diagram, while the lower-metallicity variables occupy the
higher-amplitude, shorter-period part. 
Therefore, we conclude that the metal abundance formula described by
Eq.~2 of \cite{2005AcA....55...59S} suffers from a systematic bias as a function of amplitude/period.
This finding is confirmed by a demonstrably monometallic dataset: on both panels of Fig.~\ref{fig:feh}, empty and filled circles illustrate
the $I$-band photometric metallicities of the Oosterhoff~I and II variables in the globular cluster M3,
which is well known to be monometallic \citep{1992AJ....104..645K, 2005AJ....129..303C}, highlighting
the systematic offset between these two groups of objects as well.

The second formula given by \cite{2005AcA....55...59S} in the form of Eq.~3 gives much more consistent abundance estimates as a function
of the amplitude. However, to compare the behavior of the calculated abundances of the two Oosterhoff groups of variables, we have to
separate them. This has been done using
the period-amplitude diagram, utilizing a cut that is dependent on the calculated iron abundance, as illustrated by Figure~\ref{fig:oo}.
Comparing the top and middle panels of Figure~\ref{fig:feh2} indicates that there is still a slight trend for the
Oosterhoff~I variables as a function of the amplitude, as well as an offset with respect to
the Oo~II stars, which have a calculated average [Fe/H] of $-1.05$\,dex across the whole amplitude range.
We correct for this offset using the ridge of Oo~I RRLs, derived from a KDE of amplitude
bins of the estimated abundances of Oo~I stars. The resulting rectified Oo~I metallicity estimates are consistent with the Oo~II
variables, as well as across the whole range of RRL amplitudes (bottom panel of Figure~\ref{fig:feh2}).

We do note that not only the iron abundance formulae of \cite{2005AcA....55...59S} suffer from obvious
biases when applied on a population of monometallic RRLs, nor is the problem limited to the $I$-band.
As a stopgap measure (until new, carefully calibrated photometric abundance formulae become available)
we make available a routine on \textsf{GitHub}\footnote{\url{https://github.com/gerhajdu/pyrime}} that implements
the procedure outlined here to rectify the [Fe/H] values of Oo~I RRLs calculated from Eq.~3 of \cite{2005AcA....55...59S},
given the period, as well as the $I$-band Fourier parameters $A_1$, $A_2$ and $\phi_{31}$.

\subsection{Iron abundance estimation from the $K_\mathrm{S}$-band light-curve shapes} 
\label{subsec:feh}

To assess whether it is possible to determine the iron abundance of RRLs from the $K_\mathrm{S}$ band
light-curve shapes, we made use of the $K_\mathrm{S}$-band light-curve parameters of OGLE RRLs in the VVV fields
determined in Sect.~\ref{subsec:fits}, combined with the photometric metallicities determined in Sect.~\ref{subsec:met}.
As the RRL sample on which \cite{2005AcA....55...59S} calibrated the relation used to determine the initial
[Fe/H] of the RRLs in question only extends down to [Fe/H] $= -1.7$, using even the rectified abundance estimates below this threshold is
uncertain. Nevertheless, because the typical abundance of Oo~II globular clusters bearing RRLs is about [Fe/H]$\sim -2.2$\,dex,
we are only discarding the RRLs below this limit.
This final cut results in a final sample of 6193 RRLs.

\begin{figure}[]
\epsscale{1.22}
\plotone{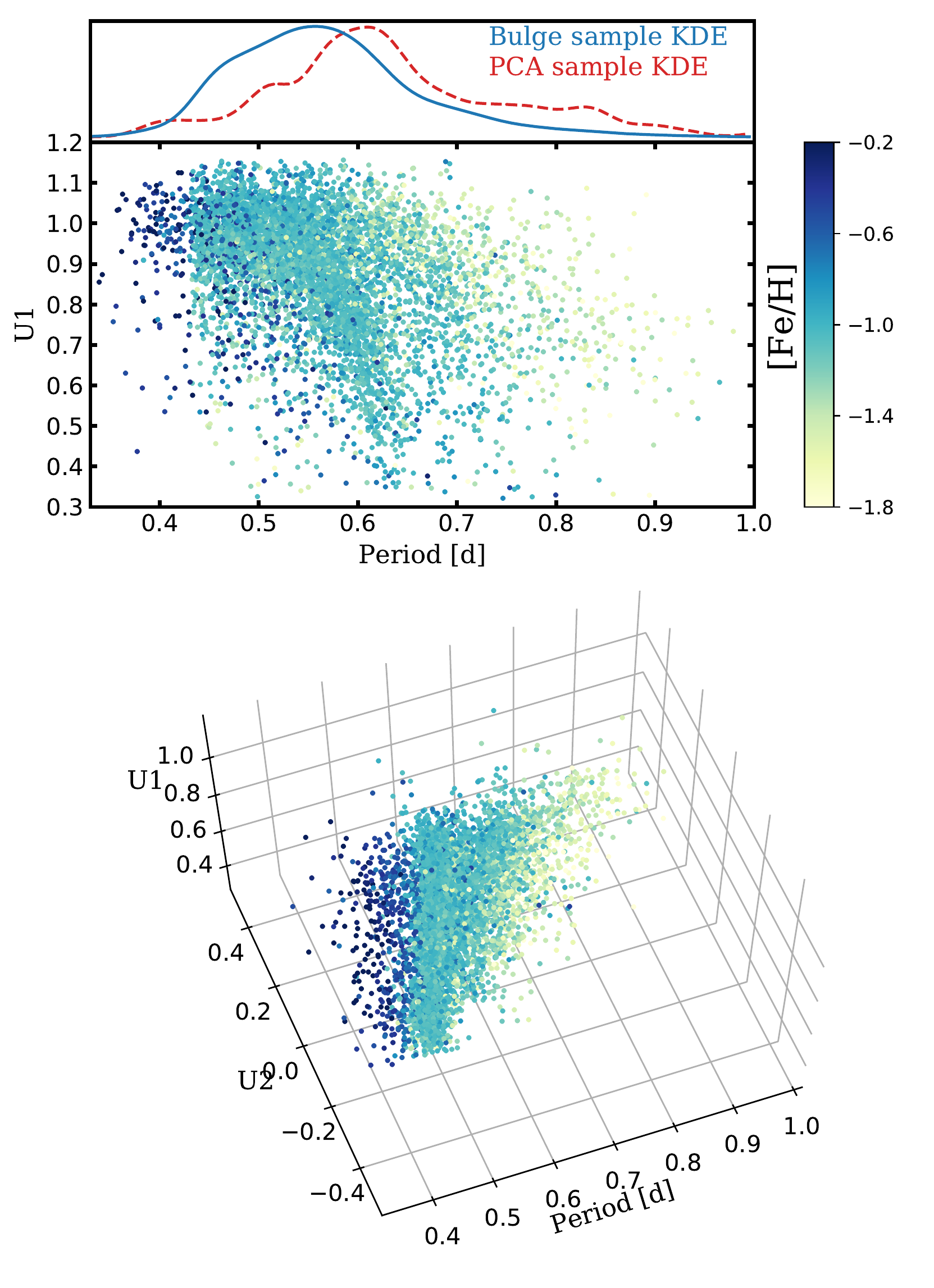}
\caption{ Dependence of the rectified photometric ($I$-band) iron abundances on the near-IR light-curve parameters.
Top: comparison of the KDE of the period distribution of the bulge (blue contiguous line)
and the PCA training sample (Table~\ref{tab:data}). Although the period distributions of the two samples are different,
the example light-curve fits of Figure~\ref{fig:examples} reveal that the method described in Sect.~\ref{sec:robust}
is well-suited for fitting the bulge RRL light curves.
Middle: the iron abundances show a significant overlap when only the period and the first PC amplitude are considered.
Bottom: at a given period and $U_1$ amplitude stars with different [Fe/H] possess different $U_2$ amplitudes,
illustrating the iron abundance dependence of the near-IR light curve shape parameters.
}
\label{fig:feh3}
\end{figure}

Figure~\ref{fig:feh3} shows the optical photometric metallicity as a function of the period, and the first two PC amplitudes.
Some general trends are apparent; for example, as expected from the form of Eq.~3 of \cite{2005AcA....55...59S}, longer-period RRLs have
higher iron abundances, but even at the same period and $U_1$, stars of different $U_2$ have systematically different iron abundances.

As additional features, we also consider Fourier parameters for this regression problem. During the light-curve fitting,
the individual PCs are represented as Fourier series (Eq.~\ref{pca_fourier}), and the total light curve is given as the
sum of these series multiplied by the PC amplitudes (Eq.~\ref{lc_model}). Therefore, traditional Fourier parameters,
such as the Fourier amplitudes and epoch-independent phase differences \citep{1981ApJ...248..291S}, can be calculated
straightforwardly. Of these, we have decided to use the amplitude of the Fourier harmonics $A_1$, $A_2$, and $A_3$,
as well as the epoch-independent phase differences $\phi_{21}$ and $\phi_{31}$.
Together with the pulsation period, the total number of independent variables is 10.

\begin{deluxetable}{c|c}
\tablenum{2}
\label{tab:hyper}
\tablecaption{Parameters of Our Hyperparameter Grid Search \label{tab:table}}
\tablehead{
\colhead{Hyperparameter} & \colhead{Candidate Parameter Value}
}
\startdata
& (20), (50), (100)  \\
Hidden layers\tablenotemark{a}& (20, 20), (20, 50), (20, 100)\\
& (50, 20), (50, 50), (50, 100)\\
& (100, 20), (100, 50), (100, 100)\\
\hline
& $10^{1.0}$, $10^{0.5}$, $10^{0.0}$, $10^{-0.5}$, $10^{-1.0}$  \\
$\alpha$\tablenotemark{b} &  $10^{-1.5}$, $10^{-2.0}$, $10^{-2.5}$, $10^{-3.0}$\\
&  $10^{-3.5}$, $10^{-4.0}$, $10^{-4.5}$, $10^{-5.0}$ \\
\hline
Activation function\tablenotemark{c} & \textsf{logistic}, \textsf{tanh}, \textsf{relu}  \\
\hline
Solver\tablenotemark{c} & \textsf{lbfgs}, \textsf{sgd}, \textsf{adam}  \\
\hline
 & $P$, $U_{1..4}$  \\
 & $P$, $U_{1..3}$, $\phi_{31}$  \\
Independent variables & $P$, $U_{1..3}$, $\phi_{21}$, $\phi_{31}$  \\
 & $P$, $A_{1..3}$, $\phi_{21}$, $\phi_{31}$  \\
& $P$, $U_{1..4}$, $A_{1..3}$, $\phi_{21}$, $\phi_{31}$  \\
\enddata
\tablenotetext{a}{Number of neurons of hidden layers of the neural network}
\tablenotetext{b}{Regularization parameter}
\tablenotetext{c}{Activation function and solver of optimization; see the \textsf{scikit-learn}
documentation for description at \url{http://scikit-learn.org/stable/modules/generated/sklearn.neural_network.MLPRegressor.html}}
\end{deluxetable}

We explored the regression routines implemented in \textsf{scikit-learn} \citep{scikit-learn} with the aim of finding a relation
to determine the iron abundance from some combination of the considered parameters. We found that regardless of the parameters,
the underlying relation is nonlinear and multimodal due to the systematic offset between the Oo I and II variables.
Additionally, the heavy excess of bulge RRLs with metallicities around [Fe/H]$=-1$\,dex biases any kind of regression without
resampling or weighting the input data.

The best results were achieved using the \mbox{\textsf{MLPRegressor}} routine, implementing
a multilayer perceptron regressor, a type of neural network. Artificial neural networks are inspired by and modeled after
biological neural networks, and have been used in many branches of science as well as in commercial applications for their ability
to learn highly nonlinear relations inherent in many types of data (see \citealt{haykin2011} for a general description of neural
networks).

Neural networks have many different parameters that have to be decided (hyperparameters) before the training of the network.
As these can vastly influence their performance, we decided to train the \mbox{\textsf{MLPRegressor}} with a grid of different
hyperparameter values, and use cross validation to determine the best combination of hyperparameters. The grid of considered hyperparameter values
is documented in Table~\ref{tab:hyper}. 
This grid only contains possible neural network architectures up to two hidden layers of neurons; although adding more hidden layers
increases the flexibility of the neural network, doing so did not significantly improve the performance in our tests.
Besides the parameters of the network, we considered different combinations of the 10 independent variables as an
additional hyperparameter.

Due to the heavily biased nature of the bulge photometric metallicity distribution (most stars have abundances [Fe/H]$\sim -1$), we implemented
a special sampling method. We grouped observations in ten 0.2\,dex wide bins (Table~\ref{tab:bins}). Because there are
only a few variables with [Fe/H] above 0 and below $-2$\,dex, these were merged with neighboring bins.
In each bin, 10 stars were selected
randomly to be part of the validation sample. Of the remaining stars, 80 variables were selected randomly with replacement to be part of the
training sample, resulting in a training set of 800 data points. As there are less than 90 stars in the most metal-poor and metal-rich bins, some
variables at the extremes of the [Fe/H] distributions are selected multiple times due to the selection with replacement.
The variables that were not selected for training were
added to the cross-validation sample. In case of an uneven distribution of stars between the two extremes in a metallicity bin,
the bin was split into sub-bins to guarantee a more even sampling as a function of metallicity.

This separation into training and cross-validation samples guarantees that the training set is large enough for the training of neural networks;
the training set is balanced on the total range of possible abundances;
by repeating this separation, variables in bins with few stars get selected for the
cross-validation sample at least a few times.

\begin{deluxetable}{c|c}
\tablenum{3}
\label{tab:bins}
\tablecaption{Number of RR~Lyrae Variables in Different Metallicity Bins}
\tablehead{
\colhead{[Fe/H] Bin} & \colhead{Number of RRLs}
}
\startdata
$[-0.2,+0.2]$ & 47\\
$[-0.4,-0.2]$ & 102\\
$[-0.6,-0.4]$ & 225\\
$[-0.8,-0.6]$ & 295\\
$[-1.0,-0.8]$ & 487\\
$[-1.2,-1.0]$ & 3888\\
$[-1.4,-1.2]$ & 628\\
$[-1.6,-1.4]$ & 375\\
$[-1.8,-1.6]$ & 84\\
$[-2.2,-1.8]$ & 62\\
\enddata
\end{deluxetable}

This separation was repeated 40 times, resulting in 40 training and cross-validation samples. The neural networks were trained
on all 40 training samples with all combinations of the possible hyperparameters detailed in Table~\ref{tab:hyper}.
In all cases, the predictions of the trained neural networks were calculated for the corresponding cross-validation samples.
Finally, for each star and for each hyperparameter combination, the predictions were averaged over the 40 repeats.
Every star appeared at least six times in the cross-validation samples.

\begin{figure}[]
\epsscale{1.10}
\plotone{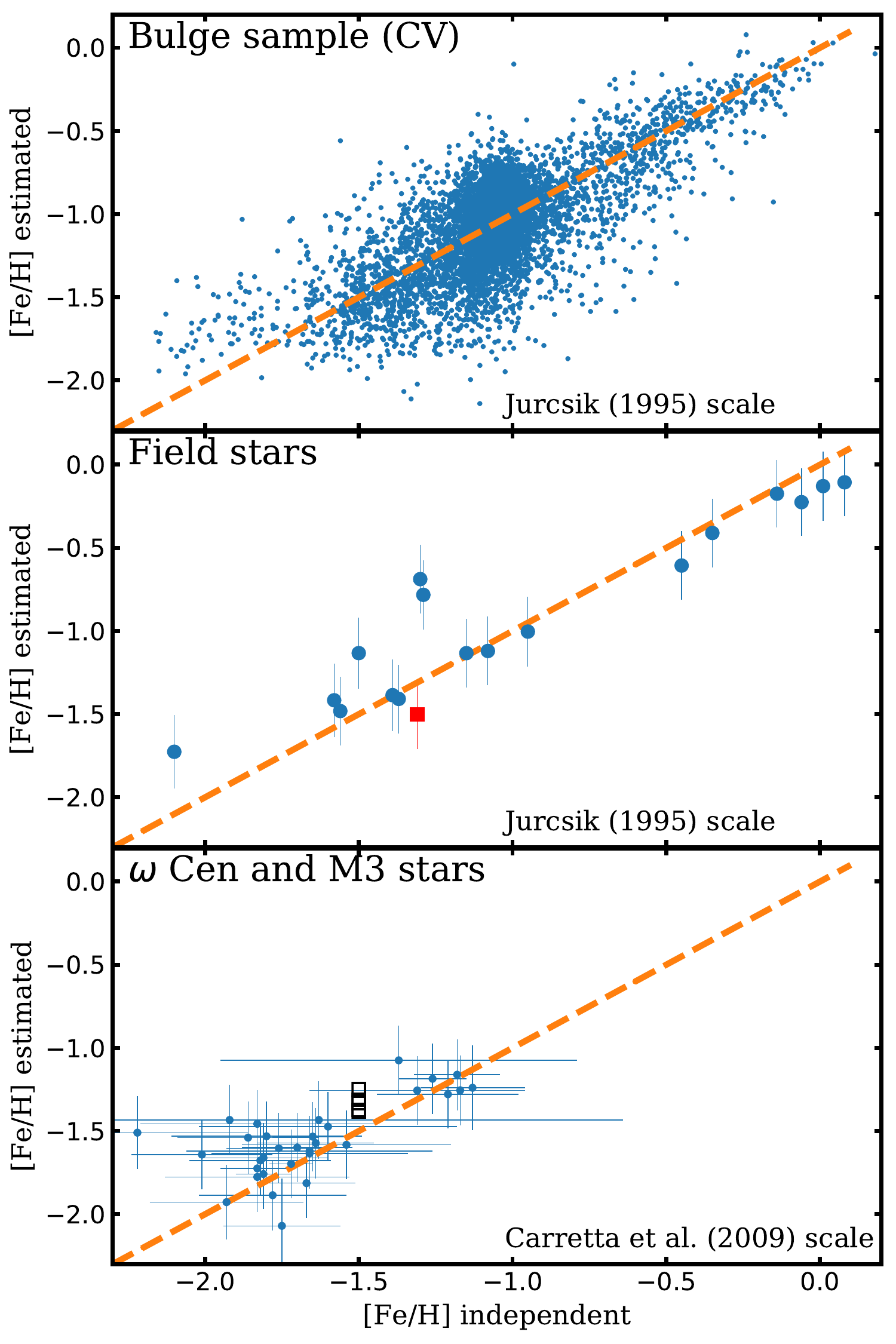}
\caption{ 
Top: cross-validated (CV)
average $K_\mathrm{S}$-band abundance estimate for 6193 RRLs in the OGLE/VVV sample
against the rectified $I$-band photometric abundance estimates (Section~\ref{subsec:met}). The dashed orange curve
represents equality between the compared abundance estimates.
Middle: comparison of estimated photometric abundances of field RRLs (circles) with their spectrophotometric
($\Delta S$-based) abundances from Table~1 of \cite{1996AA...312..111J} and the star NSV~660 (red square),
the field variable with a spectroscopic SEGUE metallicity estimate \citep{2011AJ....141...90L, 2014ApJ...780...92S}.
Individual errors are estimated as the quadratic sum of the scatter of the 100 abundance estimates
(Sect.~\ref{subsec:feh}) and a fixed uncertainty term of 0.2\,dex.
Bottom: comparison of the estimated photometric abundances of stars in the clusters $\omega$~Cen (blue dots)
and M3 (black open squares) with the individual spectroscopic measurements \citep{2006ApJ...640L..43S}
and the overall cluster iron abundance \citep{2009AA...508..695C}, respectively.
}
\label{fig:feh4}
\end{figure}

Due to the imbalanced data set, neural networks with the highest $R^2$ score (also called the coefficient of determination) calculated
on the complete sample of averaged predictions is heavily biased toward solutions predicting [Fe/H]$\sim-1$,
irrespective of the values of the independent parameters. Therefore, we have instead selected the best hyperparameter combination by using
the sum of the $R^2$ scores calculated separately for each of the abundance bins for the averaged cross-validation predictions. 
The cross-validation result with the highest score is illustrated by the top panel of Figure~\ref{fig:feh4},
with features $P$, $A_1$, $A_2$, $A_3$, $\phi_{21}$ and $\phi_{31}$;
hyperparameters $\alpha=10^0$; two hidden layers with 100 and 20 neurons; activation function \textsf{relu}; and solver \textsf{lbfgs}.

After determining the optimal hyperparameters, 100 new training samples were drawn by randomly selecting 80 stars with replacement
from each metallicity bin, but without withholding any stars for cross validation. The \mbox{\textsf{MLPRegressor}} was trained on each of
these samples, and the predicted metallicity is the average of these 100 trained regressors. The code estimating the iron abundance
using these neural networks from the $K_\mathrm{S}$-band light-curve parameters is available on \textsf{GitHub}\footnote{\url{https://github.com/gerhajdu/pymerlin}}.

\subsection{Validation of the metallicity estimates}

To provide an unbiased evaluation of the performance of a regression model, it is important to test the
performance on a test data set not used during the fit for either training or validation.
In this case, the
model was used to predict the iron abundance from the shape parameters determined directly from the PCA analysis
for the variables in Table~\ref{tab:data}. Then, these predictions can be compared to independent determinations of their
abundances from the literature.

The cross-validation results in the top panel of Fig~\ref{fig:feh4} indicate that the method described in Sect.~\ref{subsec:feh}
gives reasonable photometric abundance estimates with a scatter of $\sim 0.22$\,dex in the
$-1.7 < \mathrm{[Fe/H]} < 0.0$ abundance range. The estimated iron abundance is biased for stars below $-1.7$ toward higher values,
probably caused by the combination of overestimation of the $I$-band photometric abundances, as well as contamination of the
parameter space of low-abundance stars with those of higher estimated abundances, due to the higher relative photometric errors
in low-amplitude, long-period stars (see Fig.~\ref{fig:feh3}).

The middle panel of Fig~\ref{fig:feh4} compares the photometric abundances of 17 stars from our Table~\ref{tab:data} to
the abundances listed in Table~1 of \cite{1996AA...312..111J}. Additionally, the photometric abundance of NSV~660 is
compared with the [Fe/H] = $-1.31$ value determined from the SDSS spectra of the SEGUE Stellar Parameter Pipeline
\citep{2011AJ....141...90L,2014ApJ...780...92S}. The two values for this sample of stars generally agree. We note, however,
that there is a hint that the abundances of high-metallicity RRLs are slightly underestimated. Furthermore,
the [Fe/H] of the most metal-poor star is overestimated, in concordance with the cross-validation results.
The abundances of two stars, RR~Cet and RR~Leo, are overestimated by 0.5 and 0.6\,dex, respectively. As there is a single
source of photometry for both stars, we believe that the original $K_\mathrm{S}$-band photometry of both stars is affected by
systematic trends on their rising branches, causing the estimated light-curve shapes to be deformed, resulting in the
overestimation of their abundances.

The bottom panel of Fig~\ref{fig:feh4} compares the $K_\mathrm{S}$-band photometric abundances of the RRLs in $\omega$~Cen with their spectroscopic
measurements by \cite{2006ApJ...640L..43S}. The photometric abundances for five variables from the globular cluster
M3 are also compared to the average cluster abundance, [Fe/H] $=-1.5$\,dex, given by \cite{2009AA...508..695C}.
We do note that so far all abundances were on the metallicity scale defined by \cite{1995AcA....45..653J}.
To compare the $K_\mathrm{S}$-band photometric abundance estimates of the $\omega$~Cen and M3 variables
in a consistent way to their spectroscopic determinations, we convert them to the metallicity scale established by
\cite{2009AA...508..695C}. The conversion between the two scales can be determined by comparing the common clusters with measured abundances in
Table~1 of \citeauthor{1995AcA....45..653J} (\citeyear{1995AcA....45..653J}, J95) and Table~A1 of \citeauthor{2009AA...508..695C} (\citeyear{2009AA...508..695C}, C09)
(neglecting the cluster NGC~5927, where the difference
between the two sources is almost 0.8\,dex):

\begin{equation}
\label{eq:carretta}
{\rm [Fe/H]}_{\mathrm{C}09} = 1.044 {\rm [Fe/H]}_{\mathrm{J}95} - 0.037 ,
\end{equation}

\noindent with a residual scatter of 0.1\,dex. Furthermore, the iron abundances of \cite{2006ApJ...640L..43S} are also converted
to the \cite{2009AA...508..695C} scale by shifting them by $-0.02$\,dex, following \cite{2016AJ....152..170B}.

The photometric
abundances of $\omega$~Cen stars reproduce the bimodal distribution of iron abundances \citep{2006ApJ...640L..43S}.
When taking into account the uncertainties of individual
measurements made by \cite{2006ApJ...640L..43S}, there is no offset for the metal-rich group at $-1.2$\,dex.
For the metal-poor group of variables, there is a hint that our method systematically overestimates abundances
by about $0.1$\,dex, similarly to the results on the top and middle panels of the Figure~\ref{fig:feh4}. As for the M3 RRLs,
their average $K_\mathrm{S}$-band photometric abundance estimate is [Fe/H] $=-1.33$\,dex, only $0.17$\,dex higher
than the cluster metallicity given by \cite{2009AA...508..695C}.

During the calibration of the abundance prediction, we discarded Blazhko variables, as their modulated light curves lead to
systematic biases in their calculated photometric abundances (see, e.g., Fig~4 of \citealt{1996AA...312..111J}). Nevertheless,
a significant fraction of the total population of RRLs suffers from this light-curve modulation. 
The nature of the Blazhko effect in the near-IR has not been established until very recently,
due to the lack of well-populated light curves covering sufficiently long time spans.
In \cite{jurcsik2018}, we show that the modulation is indeed present in the $K_\mathrm{S}$-band light curves, but with diminished
amplitudes. We have no abundance estimates for the 22 Blazhko RRLs studied in \cite{jurcsik2018}, but we surmise that the majority
of them must come from the bulge population of RRLs. We calculated the abundances based on their individual $K_\mathrm{S}$-band mean
light curves, for which we get a mean of [Fe/H]$=-1.11$\,dex with a standard deviation of $0.28$\,dex. These values are in agreement
with those of the original bulge sample, where the mean and standard deviation are [Fe/H]$=-1.06$\,dex and $0.26$\,dex, respectively.
Therefore, we surmise that the iron abundances of RRLs should not show systematic biases when estimated with the parameters
of the complete, long-term near-IR light curves, even when the Blazhko effect is present.

In summary, the $K_\mathrm{S}$-band light-curve parameters allow the estimation of the iron abundance of RRLs to an accuracy
of $0.20 - 0.25$\,dex, with slight hints of systematic trends (at the $<0.05$\,dex level) in the range of $-1.7 < \mathrm{[Fe/H]} < 0$
(underestimation for high abundances, overestimation for low ones), but abundances lower than this range are systematically overestimated.
To establish better relations, one will need high-quality $K_\mathrm{S}$-band observations of a large sample of
RRLs in globular clusters covering a wide metallicity range, and/or $K_\mathrm{S}$-band observations of the hundreds of bright
RRLs in the Solar neighborhood, and/or spectroscopic iron abundances for the OGLE field RRLs, with emphasis on the
high and low-abundance extremes.

\section{Summary} \label{sec:discussion}

In this study, we have analyzed the near-IR light-curve properties of RRab variables. The principal component analysis of
the $K_\mathrm{S}$-band light curves of a sample of 101 RRLs revealed that their varied shapes can be described
as a linear combination of a low number of PCs. It has to be stressed that compared with most previous studies
involving PCA of light curves, we decided to phase our variables by the light-curve minimum, as the near-IR maxima of most
RRLs is nearly flat and hard to localize. We advocate the 
exploration of alternatives to phasing the light curves of pulsating variables by their maxima, even for studies in the optical.
Possibilities include, among others, phasing by the light-curve minima, as was done here; by the middle of the rising or descending branches;
or by the phase of the first harmonic of a Fourier series.

The comparison of the $K_\mathrm{S}$-band light curve parameters and the $J$-band light curves of 87 variables
led to the conclusion that the former can be used to reliably predict the light-curve shape in the $J$ band. This finding is of great interest for
time-series surveys where most data points are taken in a single filter (as is the case of the VVV survey), as the accurate
approximation of the light-curve shape in a different band can greatly reduce the number of observations needed to determine accurate
mean magnitudes in complementary bands, hence truly representative colors. This can be especially useful for the
estimation of the amount of foreground reddening toward individual stars.

We developed a method to fit the RRL light-curve shapes in the $K_\mathrm{S}$ band with the PCs as basis vectors,
while also taking into account some of the peculiarities of the VVV data. The routine takes advantage of the robustness of
the Huber loss function, in order to
decrease the effect of outliers in VVV light curves. The $J$-band light-curve shapes are also approximated from the PC
amplitudes, resulting in accurate mean magnitudes in the $JHK_\mathrm{S}$ bands.

The possibility of deriving metallicities from the $K_\mathrm{S}$-band light curve parameters, analogous to what was done in
the optical \citep{1996AA...312..111J}, was explored with the help of RRLs in common between the VVV and OGLE surveys. The PC
amplitudes of these stars were determined using our fitting method on the VVV light curves. However, the photometric metallicities
of these variables, as calculated using Eq.~2 of \cite{2005AcA....55...59S} on the $I$-band light-curve parameters, revealed
a systematic trend with the period/amplitude of the main locus of variables, representing the RRL population of the
Galactic bulge. The same trend is also seen in other populations, as demonstrated by Figure~\ref{fig:feh} for the RRLs
of the monometallic globular cluster Messier~3. This problem with photometric metallicities is especially worrying in view of their prevalent
use when tracing old populations with RRLs. Therefore, we emphasize the necessity of calibration and validation of such relations
on stars spanning the whole range of possible amplitudes, periods, light-curve shapes and iron abundances of RRLs.

The relation presented by \cite{2005AcA....55...59S} in the form of Eq.~3 has a smaller, but still detectable trend as a function of
the amplitude, which we have corrected for. These rectified photometric iron abundance estimates were used to look for relations
between [Fe/H] and the $K_\mathrm{S}$-band light-curve shapes of the bulge RRL sample. As the apparent relations are nonlinear,
and the data set is unbalanced, we have decided to use the aggregate of many separate neural networks trained on balanced sub-sets of the
total dataset. The resulting accuracy for the determination of individual iron abundances based on the $K_\mathrm{S}$-band light curves
is about $0.20 - 0.25$\,dex.

The methods developed in this work are utilized in a companion paper \citep{dekany} to characterize the RRL population of
the disk area of the VVV survey, allowing the estimation of the metallicity distribution function of RRL stars in that area.
Some of these approaches could be adopted relatively easily for the exploration of other sources of time-series photometry, for
example where the small number of data points or the disparate amount of multiband observations do not lend themselves well
to traditional (i.e., Fourier based, in the case of pulsating stars) methods. Naturally, in order to utilize such
\textit{prior} knowledge, a high-quality (as well as preferably expansive) training sample is needed to characterize
the population of variable stars in question.

%
%
%
%
%

\acknowledgments

G.H. acknowledges support from the Graduate Student Exchange Fellowship
Program between the Institute of Astrophysics of the Pontificia Universidad
Cat\'olica de Chile and the Zentrum f\"ur Astronomie der Universit\"at
Heidelberg, funded by the Heidelberg Center in Santiago de Chile and the
Deutscher Akademischer Austauschdienst (DAAD), and by the
CONICYT-PCHA/Doctorado Nacional grant 2014-63140099.
G.H. and M.C. acknowledge support by the Ministry for the Economy, Development,
and Tourism's Programa Iniciativa Milenio through grant IC120009, by
Proyecto Basal PFB-06/2007, by FONDECYT grant \#1171273, and by
CONICYT's PCI program through grant DPI20140066.
M.C. gratefully acknowledges additional support by the
DAAD and the Deutsche Forschungsgemeinschaft (DFG).
I.D. and E.K.G. were supported by the Sonderforschungsbereich SFB 881
``The Milky Way System" (subproject A3) of the DFG.
Processing and analysis of data were partly performed on the Milky Way
supercomputer, which is funded by the Sonderforschungsbereich SFB 881
``The Milky Way System" (subproject Z2) of the DFG.


%

\vspace{5mm}
\facilities{ESO:VISTA}

\software{\textsf{scikit-learn} \citep{scikit-learn},  
          \textsf{scipy} \citep{scipy}
          }




\end{document}